\documentclass[12pt,a4paper,onecolumn,draftclsnofoot]{IEEEtran}

\title{Ray-of-Arrival Passing for Indirect Beam Training in Cooperative Millimeter Wave MIMO Networks}
\author{Matthew Kokshoorn, He Chen, Yonghui Li, and Branka Vucetic}

\usepackage[T1]{fontenc}
\usepackage[utf8]{inputenc}     

\usepackage{color}              
\usepackage{enumitem}
\usepackage{fancybox,fancyhdr,setspace}     
\usepackage{amssymb,amsthm}         
\usepackage[cmex10]{amsmath}
\usepackage{thmtools}                       

\usepackage{supertabular,array}             
\usepackage{booktabs}                       
\usepackage{tabularx}                       

\usepackage[margin=10pt,font=small,labelfont=bf,indention=0.75cm,labelsep=endash]{caption}
\usepackage{float}                          
\usepackage{ifthen}                         

\usepackage{subfigure}                         
\usepackage[bottom,stable]{footmisc}        

\usepackage[normalem]{ulem}                 
\usepackage{textcomp}                       
\usepackage{gensymb}                        
\usepackage[acronym,section=subsection,nomain]{glossaries} 
\makenoidxglossaries   

\usepackage{cite}

\usepackage{listings}

\usepackage{dirtree}                        

\usepackage{graphicx}
\definecolor{urlcolour}{rgb}{0,0,0.6}

\usepackage{mdframed}                      
\usepackage{tikz}

\usepackage{color}
\usepackage{bm}

\usepackage[ruled,linesnumbered]{algorithm2e}
\makeatletter
\newcommand{\removelatexerror}{\let\@latex@error\@gobble}
\makeatother
\usepackage{algorithmicx}

\usepackage{pifont}

\usepackage[utf8]{inputenc}
\usepackage{graphicx}
\usepackage{amssymb}
\usepackage{color}

\declaretheoremstyle[
spaceabove=6pt, spacebelow=6pt,
headfont=\normalfont\bfseries,
notefont=\normalfont\itshape, notebraces={}{},
bodyfont=\normalfont,
postheadspace=1em,
headindent=2.3em,
headpunct=\normalfont\itshape{:},
]{INDENTthm}

\declaretheoremstyle[
spaceabove=6pt, spacebelow=6pt,
headfont=\normalfont\bfseries,
notefont=\normalfont\itshape, notebraces={}{},
bodyfont=\normalfont,
postheadspace=1em,
headindent=-2.3em,
headpunct=\normalfont\itshape{:},
]{INDENTlem}

\makeatletter
\renewcommand\thmt@mklistcmd{%
	\@xa\protected@edef\csname l@\thmt@envname\endcsname{
		\@nx\@dottedtocline{1}{1.5em}{\@nx\thmt@listnumwidth}{\thmt@thmname}{mu}%
	}%
	\ifthmt@isstarred
	\@xa\def\csname ll@\thmt@envname\endcsname{%
		\protect\numberline{\thmt@thmname\protect\let\protect\autodot\protect:}%
		\ifx\@empty\thmt@shortoptarg\else\protect\thmtformatoptarg{\thmt@shortoptarg}\fi
	}%
	\else
	\@xa\def\csname ll@\thmt@envname\endcsname{%
		\thmt@thmname\ \csname the\thmt@envname\endcsname:\hfil%
		\ifx\@empty\thmt@shortoptarg\else\thmt@shortoptarg\fi
	}%
	\fi
	\@xa\gdef\csname thmt@contentsline@\thmt@envname\endcsname{%
		\thmt@contentslineShow
	}%
}
\makeatother

\declaretheorem[style=INDENTthm,title=Theorem]{thm}

\declaretheorem[style=INDENTlem,title={\hspace{2.3em}Lemma},within=thm]{lemma}

\declaretheorem[style=INDENTlem,title={\hspace{2.3em}Corollary},within=thm]{colar}

\declaretheorem[style=INDENTlem,title={\hspace{2.3em}Example},within=thm]{exth}

\declaretheorem[style=INDENTlem,title={\hspace{2.3em}Note},within=thm]{note}

\theoremstyle{remark}

\newenvironment{myindentpar}[1]%
{\begin{list}{}%
		{\setlength{\leftmargin}{#1}\setlength{\rightmargin}{#1}}%
		\item[]%
	}
{\end{list}}

\usepackage[subfigure]{tocloft}

\usepackage{glossary-mcols}

\usepackage{rotating}                       

\usepackage[final]{pdfpages}                

\definecolor{highlightcolor}{rgb}{1,0,0}    

\definecolor{todocolour}{rgb}{1,0.3,0.2}
\definecolor{TODOcolour}{rgb}{1,0,0}
\newcommand{\todo}[2][brackets]{\textsf{\textcolor{todocolour}{%
    \ifthenelse{\equal{#1}{brackets}}{[TODO: #2]}{TODO: #2}%
}}}

\let\oldmarginpar\marginpar
\renewcommand\marginpar[1]{\-\oldmarginpar[\raggedleft\scriptsize #1]%
{\raggedright\scriptsize #1}}

\def\dummyfigure{LaTeX/dummy}%

\newcommand{\incgfx}[2]{%
    \def\figfilename{\dummyfigure}%
    \def\testfile{\chapdir/Figures/#2}%
    \IfFileExists{\testfile.jpg}{\def\figfilename{\testfile}}{}%
    \IfFileExists{\testfile.png}{\def\figfilename{\testfile}}{}%
    \IfFileExists{\testfile.pdf}{\def\figfilename{\testfile}}{}%
    \IfFileExists{\testfile.jpeg}{\def\figfilename{\testfile}}{}%
    \IfFileExists{\testfile.tif}{\def\figfilename{\testfile}}{}%
    \IfFileExists{\testfile.tiff}{\def\figfilename{\testfile}}{}%
    \includegraphics[#1]{\figfilename}%
}%

\definecolor{examplebackground}{cmyk}{0, 0.005, 0.06, 0.03}
\newlength{\exmargin}   
\newlength{\exlinewidth}
\newtheoremstyle{thexample}
    {\topsep}
    {\topsep}
    {}
    {}
    {\bfseries}
    {}
    {0pt}
    {\thmname{#1}\thmnumber{ #2}\thmnote{ --- #3}}

\def\thbegin{\begin}
    {\thbegin{mdframed}
    [linewidth={\exlinewidth},leftmargin={\exmargin},rightmargin={\exmargin},backgroundcolor={examplebackground}]
    \thbegin{thexample}[#1]#2\hspace{0pt}\nopagebreak
    \setstretch{1.15}
    \nopagebreak}
    {\end{thexample}\end{mdframed}}%
\makeatletter
\renewcommand{\l@thexample}[2]{\@dottedtocline{1}{1.5em}{2.3em}{#1}{#2}\vspace{-1.3\parskip}}
\makeatother

\newacronym[description={Millimeter Wave}]
			{mmWave}{mmWave}{millimeter wave}	

\newacronym[description={Analog-to-Digital Converter},
			plural={ADCs},
			longplural={Analog-to-Digital Converters}]
			{ADC}{ADC}{analog-to-digital converter}

\newacronym[description={Digital-to-Analog Converter},
			plural={DACs},
			longplural={digital-to-analog converters}]
			{DAC}{DAC}{digital-to-analog converter}

\newacronym[description={Signal-to-Noise Ratio},
			plural={SNRs},
			longplural={signal-to-noise ratio}]
			{SNR}{SNR}{signal-to-noise ratio}

\newacronym[description={Base Station},
			plural={BSs},
			longplural={base stations}]
			{BS}{BS}{base station}
			
\newacronym[description={User Equipment},
			plural={UE},
			longplural={user equipment}]
			{UE}{UE}{user equipment}

\newacronym[description={Probability of Estimation Error}]
			{PEE}{PEE}{probability of estimation error}
			
\newacronym[description={Multiple-Input Multiple-Output}]
			{MIMO}{MIMO}{multiple-input multiple-output}

\newacronym[description={Uniform Linear Array},
			plural={ULAs},
			longplural={uniform linear arrays}]
			{ULA}{ULA}{uniform linear array}

\newacronym[description={Linear Minimum Mean Squared Estimator}]
			{LMMSE}{LMMSE}{linear minimum mean squared estimator}

\newacronym[description={Maximum Likelihood}]
			{ML}{ML}{maximum likelihood}
			
\newacronym[description={Compressed Sensing}]
			{CS}{CS}{compressed sensing}

\newacronym[description={Random Directional Beamforming}]
			{RDB}{RDB}{random directional beamforming}
			
\newacronym[description={Exhaustive Search}]
			{ES}{ES}{Exhaustive Search}

\newacronym[description={Additive White Gaussian Noise}]
			{AWGN}{AWGN}{additive white Gaussian noise}
			
\newacronym[description={Radio Frequency}]
			{RF}{RF}{radio frequency}

\newacronym[description={Angle-of-Departure},
			plural={AoDs},
			longplural={angles-of-departure}]
			{AOD}{AoD}{angle-of-departure}

\newacronym[description={Angle-of-Arrival},
			plural={AoAs},
			longplural={angles-of-arrival}]
			{AOA}{AoA}{angle-of-arrival}

\newacronym[description={Ray-of-Departure},
			plural={RODs},
			longplural={rays-of-departure}]
			{ROD}{RoD}{ray-of-reparture}
			
\newacronym[description={Ray-of-Arrival},
			plural={ROAs},
			longplural={rays-of-arrival}]
			{ROA}{RoA}{ray-of-arrival}

\newacronym[description={Analog Fountain Code}]
			{AFC}{AFC}{Analog Fountain Code}

\newacronym[description={Generalized Approximate Message Passing}]
			{GAMP}{GAMP}{Generalized Approximate Message Passing}
			
\newacronym[description={Least Absolute Shrinkage and Selection Operator}]
			{LASSO}{LASSO}{Least Absolute Shrinkage and Selection Operator}

\newacronym[description={Bernoulli Gaussian}]
			{BG}{BG}{Bernoulli Gaussian}    

\newacronym[description={Fast Channel Estimation}]
			{FCE}{FCE}{Fast Channel Estimation}    
			
\newacronym[description={Simultaneous estimation With Iterative Fountain Training}]
			{SWIFT}{SWIFT}{Simultaneous estimation With Iterative Fountain Training}
			
\newacronym[description={Forcing Probability Adaptation}]
			{FPA}{FPA}{Forcing Probability Adaptation}
			
\newacronym[description={Partially Estimated Probability Adaptation}]
			{PEPA}{PEPA}{Partially Estimated Probability Adaptation}
			
\newacronym[description={Ray-of-Arrival Passing for In-Direct}]
			{RAPID}{RAPID}{Ray-of-Arrival Passing for In-Direct}
			
\newacronym[description={Rate Adaptive Channel Estimation}]
			{RACE}{RACE}{rate adaptive channel estimation}

\newacronym[description={Probability Density Function}]
			{PDF}{PDF}{probability density function}

\newacronym[description={Cumulative Density Function},
			plural={CDFs},
			longplural={cumulative density functions}]
			{CDF}{CDF}{cumulative density function}

\newacronym[description={Circularly Symmetric Complex Gaussian}]
			{CSCG}{CSCG}{circularly symmetric complex Gaussian}

\newacronym[description={Channel State Information}]
			{CSI}{CSI}{channel state information}

\newacronym[description={Line-of-Sight}]
			{LOS}{LOS}{line-of-sight}
\newacronym[description={Non-Line-of-Sight}]
			{NLOS}{NLOS}{non-line-of-sight}

\newacronym[description={Code Division Multiple Access}]
			{CDMA}{CDMA}{Code Division Multiple Access}
\newacronym[description={Time Division Multiple Access}]
			{TDMA}{TDMA}{Time Division Multiple Access}

\begin{document}

\maketitle

\begin{abstract}
	This paper is concerned with the channel estimation problem in multi-cell \gls{mmWave} wireless systems. We develop a novel \gls{RAPID} framework, in which a network consisting of multiple BS are able to work cooperatively to estimate jointly the UE channels. To achieve this aim, we consider the spatial geometry of the \gls{mmWave} environment and transform conventional angular domain beamforming concepts into the more euclidean, Ray-based domain. Leveraging this model, we then consider the conditional probabilities that pilot signals are received in each direction, given that the deployment of each BS is known to the network. Simulation results show that \gls{RAPID} is able to improve the average estimation of the network and significantly increase the rate of poorer quality links. Furthermore, we also show that, when a coverage rate threshold is considered, RAPID is able to improve greatly the probability that multiple link options will be available to a user at any given time. 
\end{abstract}

\section{Introduction}

In order to meet the unprecedented throughput demands of next-generation communications systems, mobile networks are expected to become significantly denser in urban areas \cite{ge20165g,ultradense2015,userCentric2016}. By reducing the typical cell size, the number of devices that each \gls{BS} needs to support can be decreased; however, the smaller inter-cell spacing can lead to increased interference between cells \cite{GuizaniDense2016}. Supporting densification, the \gls{mmWave} frequency range (30 GHz-300 GHz) is an appealing spectrum band, due to its much higher atmospheric losses. This propagation characteristic naturally attenuates inter-cell interference and thus can permit carrier frequencies to be reused in cells in closer proximity to one another \cite{bai2014coverage,kulkarni2014coverage,coverageTCOM}. Furthermore, due to its vastly underutilized spectrum, the \gls{mmWave} frequency range also offers a substantial increase in bandwidth compared to the over-congested microwave spectrum used in existing wireless systems \cite{rappaport2013millimeter,dehos2014millimeter}. Although frequency reuse may see a benefit from these propagation losses, the signals also experience significant reflection and penetration losses that together make wireless communication in the \gls{mmWave} band very challenging \cite{rappaportMeasure}. Overcoming these losses is a critical issue that must be resolved, in order for \gls{mmWave} networks to meet reliability expectations of emerging technologies such as vehicular communications and the industrial Internet-of-Things (IoT) \cite{PopovskiReliable2014,Val}.

The most widely accepted means of overcoming \gls{mmWave} propagation losses is to employ large \gls{MIMO} antenna arrays with directional beamforming \cite{roh2014millimeter,rappaport2013millimeter,Rajagopal}. Beamforming is also advantageous in further reducing interference in small cell networks, as a coordinated approach in this regard can minimize unwanted signals \cite{GuizaniDense2016,ZorziCoordination2016}. Moreover, as spacing between antenna array elements is typically proportional to the carrier wavelength, large \gls{mmWave} antenna arrays can be implemented while occupying a much smaller form factor. However, the much larger bandwidth expectation in an \gls{mmWave} system also indicates that conventional digital \gls{MIMO} architectures may have an unrealistic power consumption, due to the large number of high-rate \glspl{ADC} and \glspl{DAC}.


Even for completely digital systems, estimating large \gls{MIMO} channel matrices can be a challenging problem. At \gls{mmWave} frequencies, this becomes more difficult, due to both hardware constraints and the necessity to use beamforming to overcome propagation losses during initial access and channel estimation. On the other hand, due to these propagation losses, measurements have shown that the \gls{mmWave} channel is sparse in the spatial domain \cite{rappaport2013millimeter}. Leveraging this phenomenon, each channel can be decomposed into its underlying physical parameters, including an \gls{AOA}, an \gls{AOD}, and a path coefficient, for a small number of propagation paths. Conversely, microwave frequency \gls{MIMO} channels exhibit rich scattering environments, which leads to channel models that characterize the superposition of many paths. To capture \gls{mmWave} sparsity, an alternative \gls{MIMO} matrix representation, known as the ``virtual channel matrix,'' can be formed from the set of channel gains between each beamforming direction \cite{sayeed2002deconstructing,Sayeed}. Thanks to its sparsity, it is often advantageous to estimate this matrix directly by using \gls{CS} techniques to reduce the number of required measurements \cite{rheath,beamcodebook}.

An intuitive benchmark approach to estimating the virtual channel matrix involves exhaustively searching the channel for all possible propagation paths. This may be achieved by transmitting and receiving pilot signals while sequentially adopting combinations of beamforming vectors between the transceiver to search for any paths. Improving on this point, hierarchical codebook-based estimation strategies have been proposed, in order to reduce significantly estimation overheads by applying a divide-and-conquer type of beam search \cite{rheath,beamcodebook,xiao2016hierarchical,Kokshoorn,7579573,kokshoorn2016race}. However, as progressive beam refinement converges toward a single path in each estimation process, these approaches have an overhead that is proportional to the number of users and paths. For this reason, approaches that do not adapt toward a specific user, such as \gls{ES} and \gls{RDB}, are more appropriate, as they can support the estimation of multiple user channels in the same period \cite{kokshoorn2017beam,kokshoorn2016fountain,alkhateeby2015compressed}. Furthermore, in order to realize a robust and reliable network, it may also be important for each user to maintain multiple link options to the network \cite{multicon,initacc,handover}. In particular, this multi-connectivity would be a crucial form of redundancy to support beam switching when one path direction suddenly becomes blocked. Fortunately, in the ultra-dense, user-centric cells expected in next-generation mobile systems, it follows that each \gls{UE} will typically be in the coverage of large numbers of \gls{BS} in a given communication period. Furthermore, due to the close relationship between the environment and the physical parameters that make up the \gls{mmWave} channel, we are able to infer path directions from the network's spatial geometry.

For a sparse \gls{mmWave} channel within dense multi-BS deployments, channels between each \gls{UE} and \gls{BS} can exhibit high spatial correlation. As these channels can be expressed directly as a function of the physical environment and array orientations, channel decompositions for multiple \gls{BS} may have propagation paths that correspond to a common scatterer, or direct \gls{LOS} paths to the same \gls{UE}. Leveraging the same duality between localization and estimation, joint strategies have been proposed for microwave systems in \cite{JunttiLocal2007,WoernerLocal1996,TanLocal2008}.
Extending these concepts to \gls{mmWave} systems, \cite{SayeedmmWaveLoc2014} proposed a user cooperation estimation strategy which also able to support \gls{NLOS} estimation. Although these strategies show a significant improvement over conventional estimations, they often require a precise time difference of arrival (TDoA) or phase information, and they neglect many of the hardware constraints such as hybrid beamforming and quantized phase shifters. Furthermore, as \glspl{ULA} are typically considered in \gls{mmWave} systems, the array orientation is a commonly overlooked variable. In particular, the use of \glspl{ULA} results in an angle ambiguity problem, where ``forward'' beamforming pilots/measurements are indistinguishable from the rearward directions (See \cite{kokshoorn2017beam} and Figure 4 therein).  

Motivated by the strict hardware constraints in \gls{mmWave} systems and the need to meet network reliability requirements, we aim to develop a joint channel estimation strategy that is able to utilize spatial dependencies among multiple \gls{BS}, in order to assist the network's estimation to each \gls{UE}. Specifically, in the uplink, we consider a \gls{UE} that broadcasts beamformed pilot signals which can be jointly received by multiple \gls{BS}. We also show that, if each \gls{BS} knows the relative position of its neighbors, this physical deployment information can be utilized to identify conditional geometric relationships that exist between virtual channel estimates. To facilitate this aim, we leverage ray tracing principals to transform the channel \gls{AOD}/\gls{AOA} measurements into a more Euclidean-focused \gls{ROA} and \gls{ROD}. We then show that a given pair of \gls{ROA} measurements, received from two different positions, can be considered to sample jointly a position in Euclidean space. Similarly, we leverage geometric dependencies among the \gls{ROD} to infer conditional transmit directions. We refer to the developed scheme as ``\acrfull{RAPID} beam training.''

In contrast to the existing work, by focusing on the virtual channel information, we are able to apply our approach to hardware constrained estimation. Furthermore, to provide generality and reduce computational redundancy, we also consider that each \gls{BS} is only able to share entries from its already estimated channel. As this matrix is inherently sparse, this greatly reduces the bandwidth required to share information among the network. Furthermore, by considering the virtual channel estimate, \gls{RAPID} is agnostic to how each independent estimation is carried out, and therefore can be implemented on top of existing channel estimation strategies. Results show that the proposed scheme can greatly increase the achievable rate between the transceiver, particularly for links that would have normally been quite poor.  By considering a minimum rate requirement, we also show that \gls{RAPID} is able to significantly increase the coverage probability for having a greater number of available links. 

We summarize the main contributions of this as follows:
\begin{itemize}
	\item{  
		We investigate a multi-cell user-centric \gls{mmWave} communication system, in which a \gls{UE} broadcasts pilot signals to a number of \glspl{BS}. We generalize the concept of the \gls{AOD}/\gls{AOA} beamforming to a ray-based \gls{ROD}/\gls{ROA} estimation. We apply this model to the widely used \gls{ULA}, and develop an estimator that is robust to the angle ambiguity problem. We also show that for a \gls{BS} pair with a known relative displacement, many of their virtual channel entries are mutually dependent. 
	}
	\item{
		We use the Ray-based model to develop a Bayesian estimator so that each \gls{BS} may compute the probability of a path on its beamforming directions, given the channel estimates provided by the other network \glspl{BS}. Results show a significant improvement for both the average achievable rate and network coverage when compared to conventional schemes.
	}
	\item{
		In order to reduce sharing overheads in bandwidth constrained networks, we exploit the channel sparsity and propose to use limited information exchange among \gls{BS}. To this end, we reduce the interchange to only the most dominant virtual channel entries, that exhibit a mutually dependent relationship for another BS. In addition to this, we also show that the only prior information required for \gls{RAPID} is the relative position and orientation of each \gls{BS}. In this sense, if the \gls{UE} is also aware of this relative deployment, the proposed scheme can also be applied to the downlink. By adopting multiple access scheme for downlink pilots such as \gls{CDMA}, no sharing overhead would be required in this case.
	}
\end{itemize}

\section{System Model}

\begin{figure}[!t]
	\centering
	\includegraphics[width=5.3in,trim={8cm 4.5cm 7.4cm 4.2cm},clip]{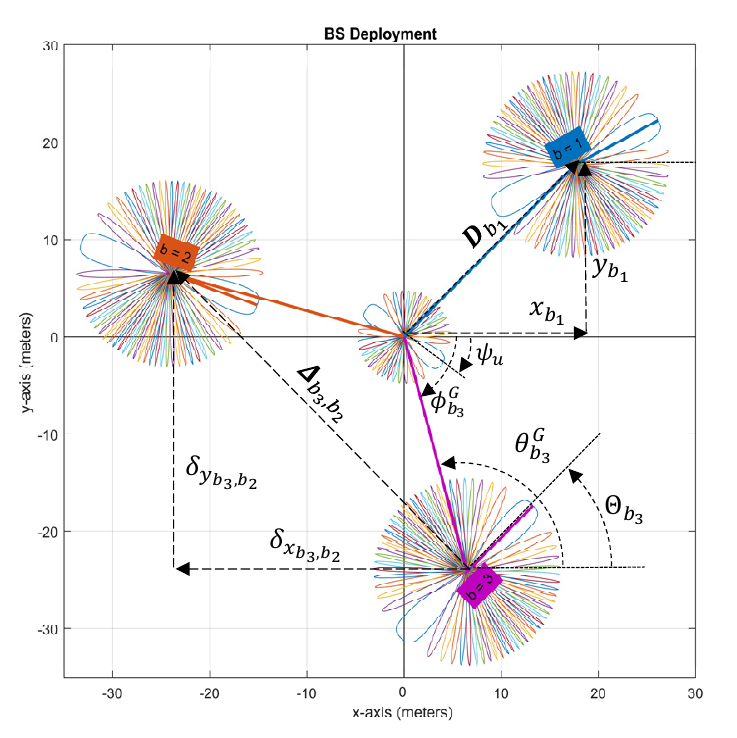}
	\caption[An example deployment illustrating each parameter in our multi-cell model.]
	{An example deployment illustrating each parameter in our model. The \gls{UE} can be seen at the origin, with antenna array orientation denoted by $\psi_u$. Similarly, each {BS} is deployed at $bm{D}_b=(x_b,y_b)$ and with orientation $\Theta_b$. To make the illustration more explicit, we show $B=3$ BS, indexed as $b_1=1$, $b_2=2$ and $b_3=3$. Relating to the Euclidean channel model, we label the \acrfull{AOD} in the {UE} local reference frame (i.e., relative to its own array orientation) ``$\phi^L_b$'' and the corresponding global reference frame (i.e., relative to the global coordinate system) angle  ``$\phi^G_b$.'' Similarly for each {BS}, we have the local reference frame \acrfull{AOA} as $\theta^L_b$ and the global angle as $\theta^G_b$.)}
	\label{RAP_BS_deploy}
\end{figure}

Consider a \gls{mmWave} cellular network consisting of $B$ \gls{BS}---each equipped with an array of $N_{\!B\!S}$ antenna. We adopt a user-centric deployment model, in which a \gls{UE} is located at the origin of a two-dimensional coordinate system (i.e., $(x_u,y_u)=(0,0)$). We further assume that the \gls{UE} is equipped with an array of $N_{\!U\!E}$ antennas. Relative to the origin of this system, we consider that the deployment of the $b$th \gls{BS} antenna array can be described by a 2D translation and a rotation, denoted by $\bm{D}_b=(x_b,y_b)$ and $\Theta_b$, respectively. We consider both the \gls{BS} and \gls{UE} antenna arrays to have an orientation denoted by $\Theta_b \in [-\pi,\pi]$ and $\psi_u \in [-\pi,\pi]$, respectively. We consider this orientation to be defined as the counter-clockwise angle from the x-axis to the \gls{ULA}. We further denote the relative displacement vector from the $p$th \gls{BS} to the $q$th \gls{BS} as $\bm{\Delta}_{p,q} = [\delta_{x_{p,q}},\delta_{y_{p,q}}] = \bm{D}_p-\bm{D}_q$ . In this paper, we use the term ``local reference frame'' to refer to angles relative to a particular antenna array. Conversely, angles in the ``global reference frame'' refer to absolute angles in the global 2D coordinate system. This distinction is important, as \gls{UE} orientation cannot be known to the network a priori. An example deployment {configuration is shown in Fig. \ref{RAP_BS_deploy}.} 

To estimate the uplink channel matrix, we assume that each \gls{UE} simultaneously broadcasts a sequence of beamformed pilot signals\footnote{Orthogonality among multiple \gls{UE} can be achieved by carrier-independent multiple access schemes such as \gls{CDMA} or \gls{TDMA}.}. Similarly, all \gls{BS} collect these signals by adopting a sequence of beamforming vectors. We consider that both the \gls{UE} and each \gls{BS} are equipped with a limited number of \gls{RF} chains, denoted by $R_{\!B\!S}$ and $R_{\!U\!E}$, respectively. Denote $\bm{f}_i$ as the $N_{\!U\!E} \times 1$ transmit beamforming vector adopted by the $i$th \gls{RF} chain at the \gls{UE}. Similarly, denote by $\bm{w}_j^{(b)}$, the $N_{\!B\!S} \times 1$ receiving beamforming vector adopted by the $j$th \gls{RF} chain of the $b$th \gls{BS}.

Following \cite{rheath}, we consider the beamforming vectors, at each link end, as being limited to networks of \gls{RF} phase shifters. As such, all elements of $\bm{f}_i$ and $\bm{w}_i^{(b)}$ are {constrained to have} a constant modulus and unit norm, such that $||\bm{f}_i||=1, \forall$ $i =1,\cdots ,R_{\!U\!E}$, and $||\bm{w}_j^{(b)}||=1, \forall$ $j =1,\cdots ,R_{\!B\!S}, b =1,\cdots ,B$. We further assume that due to hardware constraints, each of the phase shifters (i.e., the entries of $\bm{f}_i$ and $\bm{w}_j^{(b)}$ ) is digitally controlled and takes on quantized values from the predetermined set
\begin{align} \label{RAP_BS_set}
\left\{ \frac{1}{\sqrt{N}}\text{exp}(j q_k) \right\}, \forall k=1,\cdots,N,
\end{align}  
where $q_k=\pi-2\pi(k-1)/N$ and $N\in\{N_{\!U\!E},N_{\!B\!S}\}$ is the number of antennas in the array. That is, each \gls{UE} (\acrshort{BS}) phase shifter can only use one $N_{\!U\!E}$ ($N_{\!B\!S}$) uniformly spaced point around the unit circle, respectively, which can therefore be digitally controlled by $\lceil\text{log}_2N\rceil$ bits.

Let $\bm{F}=[\bm{f}_1,\bm{f}_2,\cdots,\bm{f}_{R_{\!U\!E}}]$ denote the $N_{\!U\!E}\times R_{\!U\!E}$ \gls{UE} beamforming matrix, with columns representing the $R_{\!U\!E}$ \gls{RF} beamforming vectors. The corresponding $N_{\!U\!E}\times 1$ \gls{UE} transmit signal can be represented as
\begin{align} \label{RAP_x}
\bm{x} = \sqrt{\frac{P}{R_{\!U\!E}}} \bm{F} \bm{s},
\end{align}
where $P$ is the \gls{UE}'s pilot transmit power and $\boldsymbol{s}$ is the ${R}_{\!B\!S}\times 1$ vector of transmit pilot symbols corresponding to each beamforming vector with $E[\bm{s}\bm{s}^H]=\bm{I}_{R_{\!U\!E}}$. We adopt the widely used block-fading channel model, such that the signal observed by the $b$th \gls{BS} can be expressed as \cite{alkhateeby2015compressed}
\begin{align} \label{RAP_r}
\bm{r}^{(b)} &= \bm{H}^{(b)}\bm{x} +\bm{q}^{(b)} = \sqrt{\frac{P}{R_{\!U\!E}}} \bm{H}^{(b)}\bm{F} \bm{s} + \bm{q}^{(b)},
\end{align}
where $\bm{H}^{(b)}$ denotes the $N_{\!B\!S}\times N_{\!U\!E}$ \gls{MIMO} channel matrix between the \gls{UE} and the $b$th \gls{BS}, and $\bm{q}^{(b)}$ is an $N_{\!B\!S} \times 1$ complex \gls{AWGN} vector for the $u$th user, following distribution $\mathcal{C}\mathcal{N}(\bm{0}, N_0 \bm{I}_{N_{\!B\!S}})$.

Each \gls{BS} processes the received pilot signals with each of the $R_{\!B\!S}$ \gls{RF} chains. By denoting $\bm{W}^{(b)}=[\bm{w}_1^{(b)},\bm{w}_2^{(b)},\cdots,\bm{w}_{R_{\!B\!S}}^{(b)}]$ as the $N_{\!B\!S}\times R_{\!B\!S}$ combining the matrix at the $b$th \gls{BS}, we express the $R_{\!B\!S} \times 1$ vector of the $b$th \gls{BS} received signals as
\begin{align} \label{RAP_y}
\bm{y}^{(b)} &= (\bm{W}^{(b)})^H \bm{H}^{(b)}\bm{x} +\boldsymbol{n}^{(b)}
\end{align}
where the noise term follows the distribution
\begin{align} \label{RAP_n_b_noise}
\bm{n}^{(b)}=(\bm{W}^{(b)})^H\bm{q}^{(b)} \sim \mathcal{C}\mathcal{N}(\bm{0}, N_0 (\bm{W}^{(b)})^H  \bm{W}^{(b)}). 
\end{align}

We follow \cite{Sayeed_max} and adopt a two-dimensional (2D) sparse geometric channel model. We consider that only a single dominant path is present between the \gls{UE} and each \gls{BS}, leaving the  extension to joint scatterer estimation as a future work. Using this model, each candidate uplink channel between the \gls{UE} and the $b$th \gls{BS} can be characterized in its local reference frame by an \gls{AOD}, $\phi^L_b$, an \gls{AOA}, $\theta^L_b$, and a path coefficient, namely $\alpha_b$. The corresponding \gls{MIMO} channel between the \gls{UE} and the $b$th \gls{BS} can be expressed in terms of these physical parameters as
%
\begin{align} \label{RAP_H}
\bm{H}^{(b)} = \alpha_b \sqrt{N_{\!U\!E}N_{\!B\!S}} \bm{a}_{\!B\!S}(\theta^L_b) (\bm{a}_{U\!E}(\phi^L_b))^H
\end{align}

\noindent
where $\bm{a}_{\!B\!S}(\theta^L_b)$ and $\bm{a}_{\!U\!E}(\phi^L_b)$ denote the \gls{BS} and \gls{UE} arrays' spatial signatures, respectively. We adopt a flat block fading model and assume that the path coefficient remains unchanged through the entire channel estimation process. We assume that the value of the path coefficient follows the zero mean complex distribution $\alpha_b \sim \mathcal{C}\mathcal{N}(0,\sigma_R^2)$, where the expected power, $\sigma_R^2$, is inversely proportional to the radial displacement between the \gls{BS} and \gls{UE} as $\sigma_R^2= r_b^{-\beta}$, and where $r_b=||\bm{D}_b||_2=\sqrt{x_b^2 + y_b^2}$ is the radial distance and $\beta$ is the path loss exponent.

We consider that the \gls{BS} and each \gls{UE} are equipped with \gls{ULA}. We can then write $\bm{a}_{\!B\!S}(\phi^L_b)= \bm{u}(\phi^L_b,N_{\!U\!E})$ and $\bm{a}_{\!U\!E}(\theta^L_b) = \bm{u}(\theta^L_b,N_{\!B\!S})$, respectively, whereby
%
\begin{align} \label{RAP_u}
\bm{u}(\epsilon,N) \triangleq \frac{1}{\sqrt{N}} [1,e^{j \frac{2 \pi d \text{cos}(\epsilon)}{\lambda} },\cdots,e^{j \frac{2\pi d (N-1) \text{cos}(\epsilon)}{\lambda} }]^T.
\end{align}
\noindent
In (\ref{RAP_u}), $N\in\{N_{\!U\!E},N_{\!B\!S}\}$ is the number of antenna elements in the array, $\lambda$ denotes the signal wavelength, and $d$ denotes the spacing between antenna elements. With half-wavelength spacing, the distance between antenna elements satisfies $d=\lambda/2$. 

To estimate channel information, beamforming vectors are selected from a predetermined set of candidate beamforming vectors at each link end. We denote the candidate beamforming matrices as $\bm{F}_c$ and $\bm{W}_c$, the columns of which comprise all candidate beamforming vectors at the \gls{UE} and \gls{BS}, respectively. For ease of practical implementation, we consider the candidate beams to be subject to quantized phase-shifting constraints, and therefore they represent the set of all possible beamforming vectors that may be used later for data communication. Following (\ref{RAP_BS_set}), this leads to $N_{\!U\!E}$ orthogonal transmitting candidate beams and $N_{\!B\!S}$ orthogonal receiving candidate beams. The $N_{\!B\!S} \times N_{\!U\!E}$ matrix formed by the product of the \gls{MIMO} channel and these two candidate beamforming matrices is commonly referred to as the ``virtual channel matrix'' \cite{rheath}, given by
\begin{align} \label{RAP_H_v}
\bm{V}^{(b)}=  \frac{1}{\sqrt{N_{\!U\!E}N_{\!B\!S}}} (\bm{W}_c)^H  \bm{H}^{(b)}\bm{F}_c.
\end{align}
We therefore aim to estimate this matrix so that beam pairs that result in strong channel gains can be identified for data communication. The key challenge here is determining how to design a sequence of beamforming vectors in such a way that the channel parameters can be quickly and accurately estimated, leaving more time for data communication and thus achieving a higher throughput.

To facilitate our proposed cooperative channel estimation scheme, we assume that all \gls{BS} are able to maintain a reliable link between one another and are thus able to share mutually dependent information. Initially, we consider complete information sharing, however we later restrict this to the bandwidth constrained channel by only sharing dependent measurements of significant signal strength. In the following sections, we leverage mutual channel information, in order to develop a cooperative \gls{BS} framework.  

\section{RAPID Beam Training}	

In this section, we first extend our channel model to consider the Euclidean deployment scenario. We then introduce the sequence of measurement beamforming vectors adopted in our proposed estimation scheme, and then we extend these into the 2D geometric model. {In doing so, we} propose a shift from the conventional single-link-oriented \gls{AOD} and \gls{AOA} model, to a more Euclidean-focused \gls{ROD} and \gls{ROA} model. Subsequent sections then develop a means of jointly computing the probability of each beamforming combination, given the mutually dependent information provided by cooperating \gls{BS}. 

\subsection{Euclidean Space MIMO Channel}

In order to develop our joint estimation strategy, we begin by incorporating the 2D deployment into the angular channel in (\ref{RAP_H}). To this end, we denote the {\it{global}} \gls{AOD} of the $b$th propagation path as $\phi^G_b$, i.e., the angle of the propagation from the global frame, irrespective of the \gls{UE} orientation. Similarly, we denote the global \gls{AOA} at the \gls{BS} end as $\phi^G_b$. Recalling the array orientations $\psi_u$ and $\Theta_b$, we can relate global angles into the local reference frame \gls{AOD} and \gls{AOA} (i.e., the local beam steering directions) as $\phi^L_b=\phi^G_b-\psi_u$ and $\theta^L_b=\theta^G_b-\Theta_b$. Substituting these into (\ref{RAP_H}) leads subsequently to a global description of the channel model

\begin{align} \label{RAP_H_G_1}
\bm{H}^{(b)} &= \alpha_b \sqrt{N_{\!U\!E}N_{\!B\!S}} \bm{a}_{\!B\!S}(\theta^G_b-\Theta_b) (\bm{a}_{U\!E}(\phi^G_b-\psi_u))^H.
\end{align}

By observing the geometric relationships in Fig. \ref{RAP_BS_deploy}, we can rate the global \gls{AOD} and \gls{AOA} further by considering \gls{BS} deployments with signed trigonometric relationships $\text{tan}(\phi^G_b) = y_b/x_b$ and $\text{tan}(\theta^G_b)-\pi = {y_b}/{x_b}$. Using the four-quadrant inverse tangent function, denoted by $\text{atan2}(a,b)$, the \gls{LOS} dominant channel in (\ref{RAP_H_G_1}) can be rewritten as
\begin{align} \label{RAP_H_G}
\bm{H}^{(b)} &= \alpha_b \sqrt{N_{\!U\!E}N_{\!B\!S}} \bm{a}_{\!B\!S}(\text{atan2}(-y_b,-x_b)-\Theta_b) (\bm{a}_{U\!E}(\text{atan2}(y_b,x_b)-\psi_u))^H \\
&= \alpha_b \sqrt{N_{\!U\!E}N_{\!B\!S}} \bm{E}_{\!B\!S}(-y_b,-x_b,\Theta_b) (\bm{E}_{U\!E}(y_b,x_b,\psi_u))^H
\end{align}
\begin{sloppypar}
	\noindent	
	where $\bm{E}_{\!B\!S}(-y_b,-x_b,\Theta_b) =\bm{a}_{\!B\!S}(\text{atan2}(-y_b,-x_b)-\Theta_b)$  and 
	$\bm{E}_{U\!E}(y_b,x_b, \psi_u)=$   
	$\bm{a}_{U\!E}(\text{atan2}(y_b,x_b)-\psi_u) $ describe antenna spatial signatures in terms of the Euclidean deployment parameters for the \gls{BS} and \gls{UE}, respectively. 
\end{sloppypar}

If each \gls{BS} is deployed without knowing the orientation and relative position surrounding the \gls{BS}, each {\it{independent}} estimation is limited to the angular parameters in (\ref{RAP_H_G_1}), subject to the constraints of \gls{mmWave} beamforming. Accurate estimation of these parameters permits accurate beam selection communication and consequently an accurate recovery of the fading coefficient from the pilot measurement. However, for a deployment where the orientation and relative position of surrounding \gls{BS} is known, the $b$th \gls{BS} can focus on the estimation of the parameters in (\ref{RAP_H_G}) as $\hat{x}_b$, $\hat{y}_b$ and the \gls{UE} orientation $\hat{\psi}_u$. By expressing these estimation parameters in terms of $p$th \gls{BS} as $[\hat{x}_b,\hat{y}_b] = [\hat{x}_p,\hat{y}_p]  + \bm{\Delta}_{b,p},  \forall \; b=1,\cdots,B$, it is evident that each \gls{BS} can then reconstruct not only its own channel, but also the channel of other \gls{BS} in the network. Furthermore, as the \gls{UE} can only exist in a single position, estimations among \gls{BS} are mutually dependent. This relationship supports the spatial correlation in the Euclidean channel and gives motivation for cooperation among \gls{BS} to achieve accurate joint channel estimation.

\subsection{Candidate Beamforming Measurements}

In this paper, we follow \cite{kokshoorn2017beam} and adopt random directional beam steering at each link end. To achieve this aim, we elaborate on the \gls{UE} candidate beamforming matrix in (\ref{RAP_H_v}) as $\bm{W}_{c}=[\bm{w}_c(1),...,\bm{w}_c(N_{\!U\!E})]$. Similarly, we define the \gls{BS} candidate beamforming matrix as $\bm{F}_{c}=[\bm{f}_c(1),...,\bm{f}_c(N_{\!B\!S})]$, following which, in each pilot transmission time slot, a unique pseudo-random candidate beamforming vector is adopted by each \gls{RF} chain at the \gls{UE} and similarly at each \gls{BS}. In order for each \gls{BS} to collect simultaneously and fairly pilot signals from all users, we consider the pseudo random selection of candidate beams as having equal probability\footnote{In \cite{kokshoorn2017beam}, each candidate beam is assigned a non-uniform probability of selection, which is later adaptively re-weighted to improve performance. As this results in each receiver adapting its beams toward a single user, we do not consider this approach herein.}. As each random selection is assumed to derive from a pseudo random process, the entire selection sequence can be predicted by both the \gls{UE} and each \gls{BS}, so long as the \gls{UE} maintains a synchronized random seed within the network. 

By recalling the ULA response in (\ref{RAP_u}), the resulting set of orthogonal candidate beams that satisfy the quantized phase-shifting constraints in (\ref{RAP_BS_set}) becomes \cite{kokshoorn2017beam}

\begin{align} \label{RAP_f_c}
\bm{f}_c(n_u)=\bm{a}_{U\!E}\Big( \bar{\phi}_{n_{u}} \Big), \forall \;
\bar{\phi}_{n_{u}} = \text{cos}^{-1} \Big( 1-\frac{2 \; n_u }{N_{\!U\!E}} \Big),
\; n_u\in \mathcal{N}_U
\end{align}
and
\begin{align} \label{RAP_w_c}
\bm{w}_c(n_b)=\bm{a}_{B\!S}\Big( \bar{\theta}_{n_{b}} \Big),\forall \;
\bar{\theta}_{n_{b}}=\text{cos}^{-1} \Big( 1-\frac{2 \; n_b }{N_{\!B\!S}} \Big),
\; n_b \in \mathcal{N}_B
\end{align}
where the candidate beam steering indexes at the \gls{UE} and \gls{BS} are denoted by $n_{u}th$ and $n_{b}th$, respectively, and belong to the sets $\mathcal{N}_U=\{0,\cdots,N_{\!U\!E}-1\}$ and  $\mathcal{N}_B=\{0,\cdots,N_{\!B\!S}-1\}$. Due to the quantized phase-shifting constraints, each candidate beam steering vector is orthogonal to the others, and therefore together they satisfy $ \bm{F}_{c} \bm{F}_{c}^H=\bm{F}_{c}^H \bm{F}_{c}=\bm{I}_{N_{\!U\!E}}$ and  $\bm{W}_{c} \bm{W}_{c}^H=\bm{W}_{c}^H \bm{W}_{c}=\bm{I}_{N_{\!B\!S}}$. The example set of candidate beam patterns in Fig. \ref{RAP_BS_deploy} shows each \gls{BS} with $N_{\!B\!S}=8$ and the \gls{UE} with $N_{\!U\!E}=16$. In the same figure, it is also evident that the candidate beams on the range $[0,\pi]$ are repeated in the range $[0,-\pi]$ i.e., $\bm{a}_{B\!S}(\theta)=\bm{a}_{B\!S}(-\theta)$ and $\bm{a}_{U\!E}(\phi)=\bm{a}_{U\!E}(-\phi)$, due to the one-dimensional nature of \gls{ULA}, which leads to the candidate beams' indexes ambiguously describing angles from either range. We discuss this in greater detail in subsequent sub-sections.

By using a random sequence of candidate beams to transmit and receive each pilot symbol, as described in (\ref{RAP_f_c}) and (\ref{RAP_w_c}), the sequence of $M$ measurements that are collected by the $b$th \gls{BS} can be expressed by the $R_{\!B\!S}\times 1$ measurement vector by

\begin{align} \label{RAP_y_all}
\bm{y}^{(b)} &= \sqrt{\frac{P}{R_{\!U\!E}}}
\left[\begin{array}{c}
(\bm{W}_1^{(b)})^H  \bm{H}^{(b)}\bm{F}_1 \bm{s}_1   \\
\vdots   \\
(\bm{W}_m^{(b)})^H  \bm{H}^{(b)}\bm{F}_m \bm{s}_m
\end{array}\right] +
\left[\begin{array}{c}
\bm{n}_1^{(b)} \\
\vdots   \\
\bm{n}_m^{(b)}
\end{array}\right].
\end{align}
where $\bm{F}_m$ and $\bm{W}_m^{(b)}$ are the matrices whose columns consist of the $R_{\!U\!E}$ and $R_{\!B\!S}$ randomly selected candidate beam steering vectors at the \gls{UE} and \gls{BS}, respectively. Due to orthogonality among the \gls{BS} candidate beams, the noise elements in (\ref{RAP_n_b_noise}) now follow an i.i.d., \gls{AWGN} distribution.

By rearranging (\ref{RAP_H_v}) to get $\bm{H}^{(b)} =\sqrt{N_{\!U\!E}N_{\!B\!S}} \bm{W}_c\bm{V}^{(b)}   \bm{F}_c^H$, we can substitute this result into (\ref{RAP_y_all}) and express the measurement vector in the common \gls{CS} form \cite{rangan2011generalized} as follows
\begin{align} \label{RAP_y_all_CS_1}
\bm{y}^{(b)} &=  A_g
\left[\begin{array}{c}
\bm{A}_1^{(b)} \\
\vdots   \\
\bm{A}_m^{(b)}
\end{array}\right] \text{vec}(\bm{V}^{(b)}) +
\left[\begin{array}{c}
\bm{n}_1^{(b)} \\
\vdots   \\
\bm{n}_m^{(b)}
\end{array}\right] \\
&=  A_g  \bm{A}^{(b)} \bm{v}^{(b)} + \bm{n}^{(u,m)}. \label{RAP_y_all_CS}
\end{align}
where $\bm{A}_m^{(b)}=( \bm{s}_m^T \bm{F}_m^T \bm{F}_c^*)  \otimes   (  (\bm{W}_m^{(b)})^H  \bm{W}_c  ) $ is the $R_{\!B\!S}\times N_{\!U\!E}N_{\!B\!S}$ sparse sensing matrix, $\bm{v}^{(b)}=\text{vec}(\bm{V}^{(b)})$ is the vectorized virtual channel matrix between the \gls{UE} and the $b$th \gls{BS}, and $A_g = \sqrt{P N_{\!U\!E}N_{\!B\!S}/R_{\!U\!E}}$ is a scalar measurement gain. 

\subsection{Independent Base Station Channel Estimation}

Following the measurement sequence in the previous sub-sections, we assume that each \gls{BS} independently estimates its own the virtual channel, $\hat{\bm{v}}^{(b)}$, based on measurements it has collected over the span of $T_E$ time slots in  $\bm{y}^{(b,m)}$. Considering the \gls{CS} matrix $\bm{A}^{(b,m)}$, this sparse recovery problem can formulated as
\begin{align} \label{RAP_Lasso}
\hat{\bm{v}}^{(b)} &= \underset{\bm{v}}{\operatorname{argmin}}\Big[ || \bm{y}^{(b)} -  A_g \bm{A}^{(b)} \bm{v}||_2^2 + \gamma || \bm{v} ||_1 \Big].
\end{align}

In this paper, we consider that each independent channel estimation is obtained using the \gls{BG} \gls{GAMP} approach described in \cite{vila2011expectation,Jianhua}. After obtaining this initial channel estimate, each \gls{BS} can then convert the vectorized channel estimate back into its matrix form (i.e., $\hat{\bm{V}}^{(b)}$). In the following sub-sections, we develop a framework that permits each \gls{BS} to then share its mutually dependent indexes with the rest of the network so that the joint probability of each beam combination may be computed. Although we have adopted a \gls{BG} \gls{GAMP}-based estimator in this paper, in practice \gls{RAPID} is not limited to any particular independent estimation/recovery technique; rather, any approximate solution to (\ref{RAP_Lasso}) may be considered as an input into our proposed algorithm.

\subsection{Bipolar Candidate Ray Measurements}

Following a similar process as the Euclidean channel formulation in ($\ref{RAP_H_G}$), we now also seek to transform the candidate beamforming vectors into the Euclidean deployment model. To this end, we consider that each of the candidate beamforming vectors, conventionally considered to measure an angular \gls{AOD}/\gls{AOA}, instead corresponds to a ray-based \gls{ROD}/\gls{ROA}. We model each ray to begin at the center of each \gls{ULA} and extend with a radial distance denoted by $r_b$ in the direction corresponding to each \gls{AOD}/\gls{AOA}. By adopting a bipolar parametric line model, we can describe the ($x$,$y$) coordinate pairs that lie on the $n_{b}$th candidate \gls{ROA} for the $b$th \gls{BS} as

\begin{align} \label{RAP_P_b}
\bm{P}_b = \left[\begin{array}{c} x  \\ y  \end{array} \right]  &= \left[\begin{array}{c} r_b \;\text{cos}( \bar{\theta}_{n_{b}} -\Theta_b )    \\ r_b \; \text{sin}( \bar{\theta}_{n_{b}} -\Theta_b ) \end{array} \right] + \left[\begin{array}{c} x_b  \\ y_b  \end{array} \right] \\
&= r_b \; \bm{R}(n_b) + \bm{D}_b^T, \;\; \forall \; r_b>0  \label{RAP_P_b_2}.
\end{align}
By recalling the relationship between the candidate index and angle, $\bar{\theta}_{n_{b}}  =  \text{cos}^{-1}( 1-\frac{2 \; n_b }{N_{\!B\!S}})$, we further elaborate $\bm{R}(n_b)$ as
\begin{align} \label{RAP_R_n_b}
&\bm{R}(n_b) =\left[ \!\!\! \begin{array}{c} {L}_x(n_b)  \\ L_y(n_b) \end{array} \!\! \right] =  \left[ \!\! \begin{array}{c} \text{cos}\big( \text{cos}^{-1}( 1-\frac{2 \; n_b }{N_{\!B\!S}}) -\Theta_b \big) \\ \text{cos}\big( \text{cos}^{-1}( 1-\frac{2 \; n_b }{N_{\!B\!S}}) -\Theta_b -\frac{\pi}{2} \big) \end{array} \right]  \\
&=\left[\begin{array}{c} 
\pm \sqrt{1-\big(1-\frac{2 \; n_b }{N_{\!B\!S}}\big)^2}\text{sin}(\Theta_b) +\big(1-\frac{2 \; n_b }{N_{\!B\!S}} \big) \text{cos}(\Theta_b)\\ 
\pm \sqrt{1-\big(1-\frac{2 \; n_b }{N_{\!B\!S}}\big)^2}\text{cos}(\Theta_b) -\big(1-\frac{2 \; n_b }{N_{\!B\!S}} \big) \text{sin}(\Theta_b)\\ 
\end{array} \right]. \label{RAP_R_n_b_2}
\end{align}
where the simplification in (\ref{RAP_R_n_b_2}) follows the trigonometric property 
$\text{cos}(\text{cos}^{-1}(a)-b)= \pm\sqrt{(1 - a^2)} \text{sin}(b) + a \text{cos}(b) $.


At this point, it is important to consider the square root term in $(\ref{RAP_R_n_b_2})$. In particular, it is notable that each candidate beam index corresponds to two indistinguishable \glspl{ROA} in the Euclidean space, as indicated by the plus-minus sign. As previously eluded to, this is an inherent property that arises from the use of uniform linear arrays, due to the symmetric property $\bm{a}_{B\!S}\big( +\bar{\theta}_{{n}_{b}} \big)=\bm{a}_{B\!S}\big( -\bar{\theta}_{{n}_{b}} \big)\forall \; n_b \in \mathcal{N}_B$. In the context of \gls{AOD}/\gls{AOA} estimation, this leads to an ambiguity problem, in that any given angle estimate could be one of two possibilities. For point-to-point systems, there is generally little benefit in resolving this ambiguity, as the transceiver will still be unable to direct its beam in only one of the directions\footnote{In more complex multi-user systems, this information could, however, be utilized to coordinate the reduction of interference among users \cite{ZorziCoordination2016}.}. However, in order for estimated directions to be considered in a Euclidean deployment, angle ambiguity can be an important source of uncertainty. This directional ambiguity is illustrated in Fig. \ref{RAP_RodFig}, where the \gls{UE} is shown as being positioned on two different \glspl{ROA} extending from the right-hand \gls{BS}. Although a single \gls{BS} cannot, by itself, determine which of the two \gls{ROA} directions correspond to a propagation path, there must be one globally consistent solution among all \gls{BS}. More generally, Fig. \ref{RAP_RodFig} also illustrates the \gls{ROA}-based model.

%

To consider this angle ambiguity in our proposed approach, we replace the ``unipolar'' \gls{BS} candidate beam indexes $ n_b \in \mathcal{N}_B$ with a super set of ``bipolar'' indexes $\ddot{n}_b \in \ddot{\mathcal{N}}_B =\{-\mathcal{N}_B, \; \mathcal{N}_B\}=\{-N_{\!B\!S}+1,\cdots, N_{\!B\!S}-1 \}$. With this bipolar definition, we define more rigorously each candidate beamforming vector in (\ref{RAP_w_c}) as $\bm{w}_c({n}_b)=\bm{a}_{B\!S}\big( \pm\bar{\theta}_{{n}_{b}} \big)$, which leads to
\begin{align} \label{RAP_w_c_bip}
\bm{w}_c({n}_b) = \bm{w}_c(|\ddot{n}_b|)=\bm{a}_{B\!S}\big( \bar{\theta}_{\ddot{n}_{b} } \big), \forall  
\bar{\theta}_{\ddot{n}_{b}}=
\text{sgn}(\ddot{n}_{b}) \text{cos}^{-1}  \Big( 1-\frac{2 \; |\ddot{n}_b| }{N_{\!B\!S}} \Big),
\; \ddot{n}_b \in \ddot{\mathcal{N}}_B
\end{align}
Using this definition, we can also express more explicitly the ambiguity in (\ref{RAP_R_n_b}) as
\begin{align} \label{RAP_R_sgn}
\bm{R}(\ddot{n}_b) = \left[ \!\!\! \begin{array}{c} 
\text{sgn}(\ddot{n}_b) \sqrt{1-\big(1-\frac{2 \; |\ddot{n}_b| }{N_{\!B\!S}}\big)^2}\text{sin}(\Theta_b) +\big(1-\frac{2 \; |\ddot{n}_b| }{N_{\!B\!S}} \big) \text{cos}(\Theta_b)\\ 
\text{sgn}(\ddot{n}_b)  \sqrt{1-\big(1-\frac{2 \; |\ddot{n}_b| }{N_{\!B\!S}}\big)^2}\text{cos}(\Theta_b) -\big(1-\frac{2 \; |\ddot{n}_b| }{N_{\!B\!S}} \big) \text{sin}(\Theta_b)\\ 
\end{array} \!\! \right].
\end{align}
Similarly, for each \gls{UE} candidate beamforming vector, we can elaborate on (\ref{RAP_f_c}) as
\begin{align} \label{RAP_f_c_bip}
\bm{f}_c({n}_u) = \bm{f}_c(|\ddot{n}_u|)=\bm{a}_{U\!E}\Big( \bar{\phi}_{\ddot{n}_{u} } \Big), \forall
\bar{\phi}_{\ddot{n}_{u}}=
\text{sgn}(\ddot{n}_{u}) \text{cos}^{-1}  \Big( 1-\frac{2 \; |\ddot{n}_u| }{N_{\!U\!E}} \Big),
\; \ddot{n}_u \in \ddot{\mathcal{N}}_U
\end{align}
where $\ddot{\mathcal{N}}_U =\{-\mathcal{N}_U, \; \mathcal{N}_U\}=\{-N_{\!U\!E}+1,\cdots, N_{\!U\!E}-1 \}$.

Although this bipolar \gls{ROA} model has limited benefit in a point-to-point system, it facilitates the sharing of information among transceiver arrays concurrently operating in the same Euclidean space. More specifically, when a pair of \gls{BS} adopts a pair of intersecting candidate \gls{ROA} beamforming vectors, the pair of measurements cannot be considered only to sample their independent angular directions, but also to sample the position at which the two \glspl{ROA} intercept. This increases mutual information and therefore can be used to enhance joint estimation performance. We later use geometric reasoning to find the conditional relationship between \gls{ROD} extending from the \gls{UE}.

\subsection{Identifying Mutually Dependent Rays}

We now build on the previous model by jointly considering the \gls{ROA} of another \gls{BS} operating concurrently in the same space as the $b$th \gls{BS}. Specifically, we consider the $\ddot{n}_p$th candidate \gls{ROA} extending from the $p$th \gls{BS}, where $p \neq b$. By recalling (\ref{RAP_P_b}), we can describe the common intercept between the $\ddot{n}_p$th \gls{ROA} and the $\ddot{n}_b$th \gls{ROA}, if it exists, by  
\begin{align} 
\label{RAP_P_eq}
\bm{P}_b &=\bm{P}_p \\
r_b \; \bm{R}(\ddot{n}_b) + \bm{D}_b^T &= r_p \; \bm{R}(\ddot{n}_p) + \bm{D}_p^T  
\label{RAP_P_eq_2}
\end{align}
where the solution is only valid if both radial distances are in a positive range, namely $r_b>0$ and $r_p>0$. By recalling $\bm{\Delta}_{p,b}=\bm{D}_p -\bm{D}_b =[\delta_{x_{b,p}},\delta_{y_{b,p}}]$, (\ref{RAP_P_eq}) can be rearranged to become
\begin{align} \label{RAP_P_eq_mat}
\bm{\Delta}_{p,b}^T &= \left[ \bm{R}(\ddot{n}_b)  \; \bm{R}(\ddot{n}_p)\right]  \left[\begin{array}{c} r_b \\ r_p \end{array} \right] \\
\left[ \begin{array}{c} \delta_{x_{b,p}} \\ \delta_{y_{b,p}} \end{array} \right] &= \left[\begin{array}{ccc} {R}_x(\ddot{n}_b)&   -R_x(\ddot{n}_p)& \\ {R}_y(\ddot{n}_b)&   -R_y(\ddot{n}_p)&  \end{array} \right]  \left[\begin{array}{c} r_b \\ r_p \end{array} \right]. 
\end{align}
By multiplying each side of (\ref{RAP_P_eq_mat}) by the inverse of the $2\times2$ square matrix, we can solve for the pair of radial distances from each \gls{BS} to the common intercept among their \gls{ROA}. We therefore express this pair in vector form as
\begin{align} \label{RAP_P_eq_inv}
\left[\begin{array}{c} r_b \\ r_p \end{array} \right] =\left[\begin{array}{ccc} {R}_x(\ddot{n}_b)&   -R_x(\ddot{n}_p)& \\ {R}_y(\ddot{n}_b)&   -R_y(\ddot{n}_p)&  \end{array} \right]^{-1}  \left[ \begin{array}{c} \delta_{x_{b,p}} \\ \delta_{y_{b,p}} \end{array} \right]. 
\end{align}
Leveraging the closed form expression for a $2 \times 2$ matrix inverse, we can then directly express the radial distances $r_b$ and $r_p$ as a function of the \gls{ROA} index pair $\ddot{n}_b$ and $\ddot{n}_p$ as

\begin{align} 
\label{RAP_r_b}
r_b(\ddot{n}_b,\ddot{n}_p) = \frac{ R_x(\ddot{n}_p)\delta_{y_{b,p}} -R_y(\ddot{n}_p)\delta_{x_{b,p}} }{ R_x(\ddot{n}_p){R}_y(\ddot{n}_b)-{R}_x(\ddot{n}_b)R_y(\ddot{n}_p) }, \exists \; r_b>0
\end{align}
and
\begin{align} 
\label{RAP_r_p}
r_p(\ddot{n}_b,\ddot{n}_p) = \frac{ {R}_x(\ddot{n}_b)\delta_{y_{b,p}}  -{R}_y(\ddot{n}_b)\delta_{x_{b,p}}  }{ R_x(\ddot{n}_p){R}_y(\ddot{n}_b)-{R}_x(\ddot{n}_b)R_y(\ddot{n}_p) }, \exists \; r_p>0.
\end{align}
We can then express the set of indexes that result in intercepts among the $\ddot{n}_b$th \gls{ROA} at the $b$th \gls{BS} and all \gls{ROA} at the $p$th \gls{BS} as
\begin{align} \label{RAP_r_bp_set}
\mathcal{R}_{\ddot{n}_b}^{(p)} = \Big\{
\ddot{n}_p \in \ddot{\mathcal{N}}_B \; &\Big| 
\exists \; r_b(\ddot{n}_b,\ddot{n}_p)>0 \land
r_p(\ddot{n}_b,\ddot{n}_p)>0\Big\}.
\end{align}

Finally, by noting that the solution in (\ref{RAP_r_b})-(\ref{RAP_r_p}) only depends on the relative displacement and orientation between each \gls{BS} pair, we can substitute each radial distance back into $\bm{P}_b = r_b \; \bm{R}(\ddot{n}_b) + \bm{D}_b^T$ and $\bm{D}_b^T$ to describe the corresponding set of intercept positions relative to the $b$th \gls{BS}. We therefore express the set of positions corresponding to each intercept in (\ref{RAP_r_bp_set}) as
\begin{align} \label{RAP_b_p_set}
\bm{\mathcal{P}}_{\ddot{n}_b}^{(p)} = \Big\{ \big(r_b(\ddot{n}_b, \; \ddot{n}_p)  \; {R}_x(\ddot{n}_b), r_b(\ddot{n}_b,\ddot{n}_p)  \; {R}_y(\ddot{n}_b)  \big) 
\; \ddot{n}_p \in \mathcal{R}_{\ddot{n}_b} \Big\}.
\end{align}
where each $(\tilde{x}_b, \tilde{y}_b) \in \bm{\mathcal{P}}_{\ddot{n}_b}^{(p)}$ describes the displacement, from the $b$th \gls{BS}, to an intercept between the $\ddot{n}_b$th \gls{ROA} and the $\ddot{n}_p$th \gls{ROA}.

The quantization of points described by the set in (\ref{RAP_b_p_set}) is a result of quantized phase-shifting constraints and, subsequently, the finite set of candidate beamforming directions that form the virtual channel matrix. For this reason, the grid of points formed by all intercepts from every \gls{BS} \gls{ROA} can therefore be thought of as a virtual channel in the Euclidean space. In the following sections, we develop a joint estimation tool to leverage the mutual information collected by the \gls{BS} channel measurements. Intuitively, \gls{RAPID} permits estimated information from one \gls{BS} to assist in another. 

\subsection{RAPID Beam Probabilities}

\begin{figure}[]
	\centering
	\includegraphics[width=5.30in,trim={1cm 5.0cm 0.5cm 4.2cm},clip]{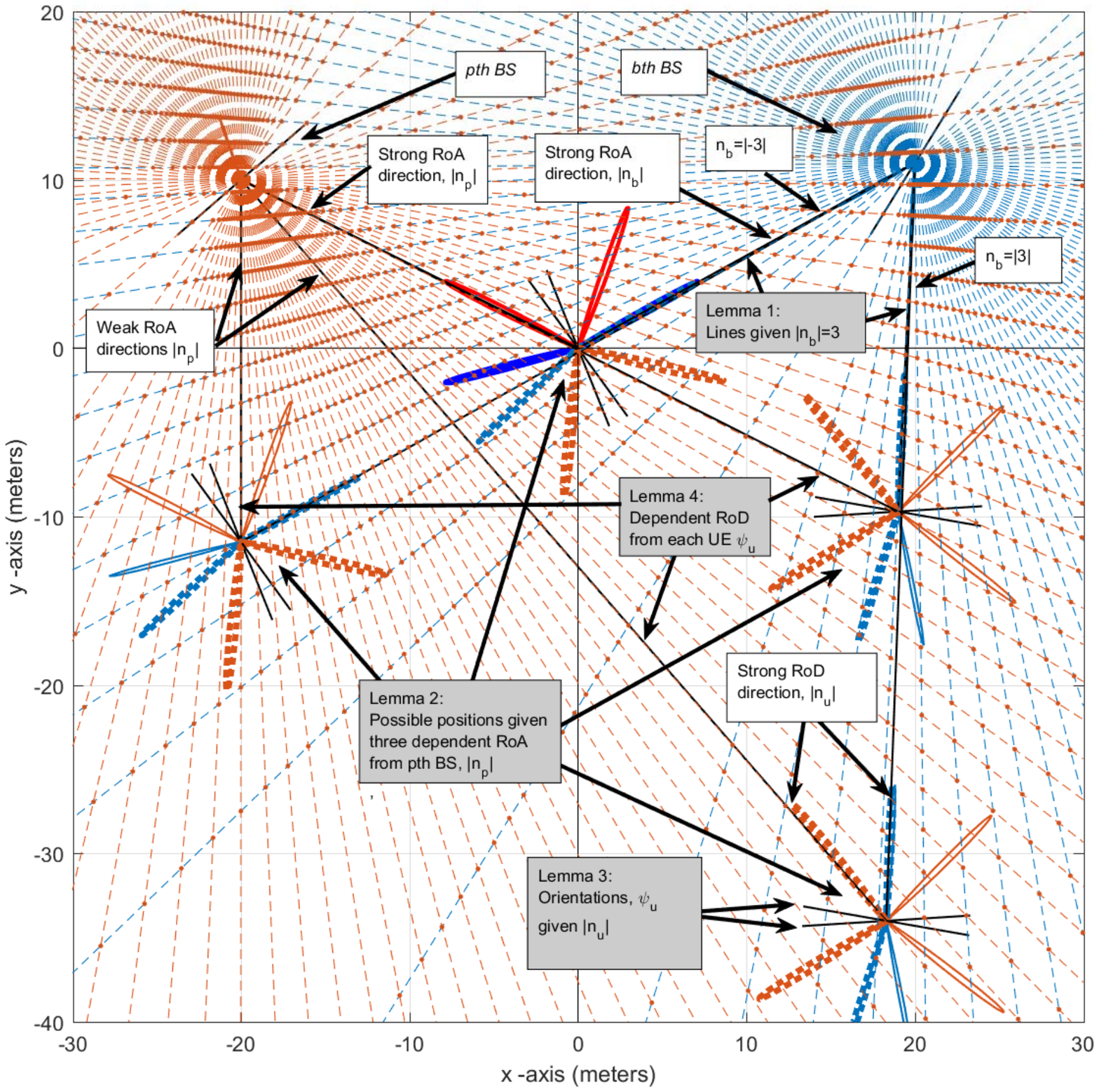}
	\caption[An example deployment model ($B=2$) showing the set of bipolar ray-based \acrfullpl{ROA} extending from the \acrfull{ULA} as dashed lines.]
	{An example deployment model ($B=2$) showing the set of bipolar ray-based
		\acrfullpl{ROA} extending from each BS as dashed lines. In this example, the true  \acrfull{UE} position is shown at the origin, with bold-solid beam patterns depicting the  \acrfull{ROD} directions expected to have strong gains. The scenario illustrated here is conditional upon the {\it{true}} $n_u$ and $n_b$. As such, the other UE positions and pairs of orientations show the conditional rotation of the correct directions. It is evident that the origin UE position aligns both $n_b$ and the conditionally dependent $n_p$. Therefore, both $\bm{V}^{(b)}_{|n_b|,|n_u|}$ and $\bm{V}^{(p)}_{|n_p|,|n_u|}$ would be expected to be strong. In each other position, either one or all of the directions do not align, and thus they correspond to expected weak codependent pairs for each virtual channel. In particular, the geometry of the lower right-hand UE position does result in the correct alignment of each \gls{AOD} and will therefore index the correct {\it{columns}} of both $\bm{V}^{(b)}$ and $\bm{V}^{(p)}$. However, as the \gls{ROA} to the $p$th BS is misaligned, the resulting joint probability will be low. Furthermore, the expected channel gain for this distant position will be significantly lower than the one observed from the true UE position at the origin.}
	\label{RAP_RodFig}
\end{figure}

From (\ref{RAP_Lasso}), recall that, after the \gls{UE} has finished transmitting its sequence of pilot symbols, each \gls{BS} is able to make an independent estimate of the up-link channel. In this sub-section, we consider that the $b$th \gls{BS} has its own estimate, denoted by $\hat{\bm{V}}^{(b)}$, and also has access to the estimated virtual channel from the $p$th \gls{BS}, denoted as $\hat{\bm{V}}^{(p)}$. Leveraging the model developed in the previous sub-sections, we now aim to utilize the mutual dependency among virtual channel entries to find the joint probability of each direction. In doing so, we therefore collectively increase overall network performance. 

We begin by considering a single entry of the $b$th \gls{BS}'s virtual channel ${\bm{V}}^{(b)}_{n_b,n_u}$, which here denotes the estimated path gain between the $n_b$th \gls{BS} candidate beamforming vector $n_u$th and the \gls{UE} candidate beamforming vector. From (\ref{RAP_f_c})-(\ref{RAP_w_c}), we can therefore elaborate this particular entry as 
\begin{align} \label{RAP_V_index_b}
{\bm{V}}^{(b)}_{n_b,n_u} = \frac{1}{\sqrt{N_{\!U\!E}N_{\!B\!S}}} ( \bm{w}_c(n_b) )^H & {\bm{H}}^{(b)} \bm{f}_c(n_u).
\end{align}
By recalling (\ref{RAP_H}), we can then consider the conditional \gls{PDF} of the channel in (\ref{RAP_V_index_b}), given the channels \gls{AOD} and \gls{AOA} that are perfectly aligned with the beamforming vectors $\bm{f}_c(n_u)$ and $\bm{w}_c(n_b)$ i.e., $\theta^L_b \in \{-\bar{\theta}_{n_{b}},\bar{\theta}_{n_{b}}\} $ and $\phi^L_b \in \{-\bar{\phi}_{n_{u}},\bar{\phi}_{n_{u}}\} $. From (\ref{RAP_H}), we can substitute these angles into (\ref{RAP_V_index_b}) to yield

\begin{align} \label{RAP_E_b_cond}
f({\bm{V}}^{(b)}_{n_b,n_u}|n_b,n_u) 
&= f(\alpha_b ( \bm{w}_c(n_b) )^H  \bm{a}_{B\!S}\big( \pm \bar{\theta}_{n_{b}} \big) 
\big(\bm{a}_{U\!E} ( \pm\bar{\phi}_{n_{u}} \big))^H
\bm{f}_c(n_u)|n_b,n_u)  \nonumber \\
&=  f(\alpha_b|n_b,n_u).
\end{align}
By itself, this conditional destiny function is the same as that used in independent estimation, with the PDF of $\hat{\alpha}_b$ being limited to $f(\alpha_b|r_b) \sim \mathcal{C}\mathcal{N}(0,r_b^{-\beta})$ for some unknown radial distance, $r_b$. However, by utilizing the models developed in the previous sub-sections, we are now able to consider (\ref{RAP_E_b_cond}) all available information and extend each conditional $n_b$ and $n_u$ into the global deployment, namely by considering the implication of each conditional for all other \gls{BS}.

By recalling array orientations and the resulting relationship between the global and local angles as $\phi^L_b=\phi^G_b-\psi_u, \forall \; b=1,\cdots, B$, and $\theta^L_b=\theta^G_b-\Theta_b, \forall \; b=1,\cdots, B$, we can deduce the following conditionally related angle sets.

\begin{thm}[\hspace{0.1cm}RAPID Theorem of Mutually Dependent Beam Angles]
	\label{RAP_RAPID_theorm}
	If the channel between the \gls{UE} and the $b$th \gls{BS} has a \gls{LOS} propagation path such that $\theta^L_b \in \{-\bar{\theta}_{n_{b}}$ and $\bar{\theta}_{n_{b}}\} $ and $\phi^L_b \in \{-\bar{\phi}_{n_{u}},\bar{\phi}_{n_{u}}\} $ with a corresponding virtual channel density function, as described by (\ref{RAP_E_b_cond}), then for the $p$th \gls{BS} to have jointly a \gls{LOS} path to the same \gls{UE}, it follows that the $p$th \gls{BS} must also have 
	\begin{align} \label{RAP_E_p_cond}
	f\left(\frac{
		( \bm{w}_c(|\ddot{n}_p|) )^H  \hat{\bm{H}}^{(p)} 
		\bm{a}_{U\!E} (\phi^L_p(\ddot{n}_b,\ddot{n}_p,n_u)   \big)
	}{\sqrt{N_{\!U\!E}N_{\!B\!S}}} \Bigg| n_b,n_u \right)
	=f(\alpha_p|n_b,n_u), \nonumber \\
	\qquad \qquad \forall \;  \ddot{n}_p \in \mathcal{R}_{\ddot{n}_b}^{(p)}
	| \exists \; \mathcal{R}_{\ddot{n}_b}^{(p)} \; \forall \;  \ddot{n}_b \in \{-n_b,n_b\}, \ddot{n}_u \in \{-n_u,n_u\}  
	\end{align}
	where the conditional \gls{AOD} to the $p$th \gls{BS}, $\phi^L_p(\ddot{n}_b,\ddot{n}_p,n_u)$, satisfies the relationship
	\begin{align} \label{RAP_phi_L_p_cond}
	\phi^L_p(\ddot{n}_b,\ddot{n}_p,n_u)  =\text{atan2}(
	r_b(\ddot{n}_b,\ddot{n}_p)  \; {R}_y(\ddot{n}_b)  -\delta_{y_{p,b}},\nonumber \\
	r_b(\ddot{n}_b,\ddot{n}_p)  \; {R}_x(\ddot{n}_b)  -\delta_{x_{p,b}})
	+ \nonumber \\ \bar{\theta}_{\ddot{n}_{b} } +\bar{\phi}_{\ddot{n}_{u} } +\Theta_b.
	\end{align}

	\begin{myindentpar}{1cm}
		\begin{lemma}[\hspace{0.1cm}Conditional Line Dependency]
			\label{RAP_RAPID_lem_a}
			In order for the $b$th \gls{BS} to receive pilot signals with the $n_b$th candidate beam, the propagation source (i.e., the \gls{UE}) must be positioned at some point along the \gls{ROA} line, indexed by $n_b = |\ddot{n}_b|, \forall \; \ddot{n}_b \in \{-n_b,n_b\}$.
		\end{lemma}

		\begin{lemma}[\hspace{0.1cm}Mutually Observable Positions]
			\label{RAP_RAPID_lem_b}
			For the $p$th \gls{BS} to have jointly received pilot signals from the same propagation source, from (\ref{RAP_r_bp_set}) it follows that there must exist some \gls{ROA} index that satisfies $\ddot{n}_p \in \mathcal{R}_{\ddot{n}_b}^{(p)}, \forall \; \ddot{n}_b \in \{-n_b,n_b\}$. That is, there must be some \gls{ROA} line extending from the $p$th \gls{BS} that intercepts with the $b$th \gls{BS}'s \gls{ROA}, indexed by $\ddot{n}_b \in \{-n_b,n_b\}$. Furthermore, this intercept has a relative displacement of $(\tilde{x}_b, \tilde{y}_b) \in \bm{\mathcal{P}}_{\ddot{n}_b}^{(p)}$ from the $b$th \gls{BS}.
		\end{lemma}
		
		\begin{lemma}[\hspace{0.1cm}Mutual Orientation Dependency]
			\label{RAP_RAPID_lem_c}
			In order for this common propagation source to qualify as being a line of sight path from the \gls{UE}, the orientation of the \gls{UE} must direct the conditionally considered $n_u$th candidate beam toward the $b$th \gls{BS}. In other words, it must satisfy the relationship $\phi^G_b=\pi-\theta^G_b$, which leads to the condition that the \gls{UE} orientation must be one of $\psi_u=\pi-\theta^L_b-\Theta_b-\phi^L_b, \forall \; \theta^L_b \in \{-\bar{\theta}_{n_{b}},\bar{\theta}_{n_{b}}\},\; \phi^L_b \in \{-\bar{\phi}_{n_{u}},\bar{\phi}_{n_{u}}\} $.
		\end{lemma}
		
		\begin{lemma}[\hspace{0.1cm}Conditional \gls{ROD}]
			\label{RAP_RAPID_lem_d}
			For the \gls{UE} to have jointly a \gls{LOS} path to the $p$th \gls{BS} (i.e., from $(\tilde{x}_b, \tilde{y}_b)$ to ($\delta_{x_{p,b}},\delta_{y_{p,b}}$), relative to the $b$th \gls{BS}), it must have a global \gls{AOD} that satisfies $\phi^G_p=\text{atan2}(\delta_{y_{p,b}}-\tilde{y}_b,\delta_{x_{p,b}}-\tilde{x}_b), \forall \; (\tilde{x}_b, \tilde{y}_b) \in \bm{\mathcal{P}}_{\ddot{n}_b}^{(p)} $. With orientation $\psi_u$ from Lemma \ref{RAP_RAPID_lem_c}, the local \gls{AOD} can then be expressed as  $\phi^L_p(\ddot{n}_b,\ddot{n}_p,n_u)=\phi^G_p -\psi_u$ and therefore as elaborated in (\ref{RAP_phi_L_p_cond}).
		\end{lemma}

		\begin{lemma}[\hspace{0.1cm}Conditional Radial Displacement and Path PDF]
			\label{RAP_RAPID_lem_e}
			If the $b$th \gls{BS} and $p$th \gls{BS} have jointly \gls{LOS} paths to a \gls{UE} positioned at the intercepting point between the \gls{ROA} pair $\ddot{n}_b$  and $\ddot{n}_p$, then, from (\ref{RAP_r_b}-\ref{RAP_r_p}), the \gls{UE} radial distance is given by $r_b(\ddot{n}_b,\ddot{n}_p)$ for the $b$th \gls{BS} and $r_p(\ddot{n}_b,\ddot{n}_p)$ for the $p$th \gls{BS}. Next, the conditional likelihood function for a $b$th \glspl{BS} path coefficient can be expressed as
			\begin{align} \label{RAP_alpha_b_cond}
			f(\hat{\alpha}_b|r_b(\ddot{n}_b,\ddot{n}_p)) &= 
			\frac{
				\text{exp} 
				\left(
				\frac{
					-|\hat{\alpha}_b|^2}{r_b(\ddot{n}_b,\ddot{n}_p)^{-\beta}+\text{Var}[\hat{\alpha}_b])} 
				\right)
			}
			{
				\pi(\; r_p(\ddot{n}_b,\ddot{n}_p)^{-\beta}+\text{Var}[\hat{\alpha}_b]))
			} 
			\end{align}
			and the path coefficient of $b$th \gls{BS} as
			\begin{align} \label{RAP_alpha_p_cond}
			f(\hat{\alpha}_p|&r_p(\ddot{n}_b,\ddot{n}_p)) = 
			\frac{
				\text{exp} \left(\frac{-|\hat{\alpha}_p|^2}{r_p(\ddot{n}_b,\ddot{n}_p)^{-\beta}+\text{Var}[\hat{\alpha}_p])} \right)
			}{
				\pi(\; r_p(\ddot{n}_b,\ddot{n}_p)^{-\beta}+\text{Var}[\hat{\alpha}_p])
			}.
			\end{align}
		\end{lemma}
	\end{myindentpar}
	\begin{colar}
		From \ref{RAP_RAPID_theorm}, we can use Bayes' rule to obtain the conditional probability of $r_b(\ddot{n}_b,\ddot{n}_p)$ as
		\begin{align} \label{RAP_Prob_b}
		Pr(&r_b(\ddot{n}_b,\ddot{n}_p)|\hat{\alpha}_b) = \frac{f(\hat{\alpha}_b|r_b(\ddot{n}_b,\ddot{n}_p))}
		{f(\hat{\alpha}_b|r_b(\ddot{n}_b,\ddot{n}_p))+f(\hat{\alpha}_b| \neg r_b(\ddot{n}_b,\ddot{n}_p))} 
		\\
		&= \frac{1}
		{1+\frac{f(\hat{\alpha}_b| \neg r_b(\ddot{n}_b,\ddot{n}_p))}
			{f(\hat{\alpha}_b|r_b(\ddot{n}_b,\ddot{n}_p))}} 
		\\ &= \frac{1}
		{1+  \big(
			\frac{ r_b(\ddot{n}_b,\ddot{n}_p)^{-\beta}}{\text{Var}[\hat{\alpha}_b]}+1 \big) 
			\text{exp}
			\left(
			\frac{ -|\hat{\alpha}_b|^2 /\text{Var}[\hat{\alpha}_b] }
			{(1 + \text{Var}[\hat{\alpha}_b]/r_b(\ddot{n}_b,\ddot{n}_p)^{-\beta}   )}
			\right)
		} 
		\end{align}
		and similarly for $Pr(r_p(\ddot{n}_b,\ddot{n}_p)|\hat{\alpha}_p)$. 
		By considering that the conditional occurrence of events $r_b(\ddot{n}_b,\ddot{n}_p)$ and $r_b(\ddot{n}_b,\ddot{n}_p)$ is, by definition, completely dependent, we can express the probability of their intersection as
		\begin{align} \label{RAP_Prob_n_given_a}
		Pr(\ddot{n}_b,\ddot{n}_p |\hat{\alpha}_b,\hat{\alpha}_p  )& =
		Pr(r_b(\ddot{n}_b,\ddot{n}_p) \cap r_p(\ddot{n}_b,\ddot{n}_p) |\hat{\alpha}_b,\hat{\alpha}_p )  \\ &= Pr(r_b(\ddot{n}_b,\ddot{n}_p)|\hat{\alpha}_b)
		Pr(r_p(\ddot{n}_b,\ddot{n}_p)|\hat{\alpha}_p) \underline{}
		\end{align}
		and the union among all mutually exclusive solutions $ \ddot{n}_p \in \mathcal{R}_{\ddot{n}_b}^{(p)}$ that are jointly conditioned with $\ddot{n}_b$ as
		\begin{align} \label{RAP_Prob_n_given_v}
		Pr(\ddot{n}_b |\hat{\alpha}_b,\hat{\alpha}_p  ) =
		Pr\Big( \;\underset{\ddot{n}_p \in \mathcal{R}_{\ddot{n}_b}^{(p)}}{\bigcup} \ddot{n}_b,\ddot{n}_p |\hat{\alpha}_b,\hat{\alpha}_p  \Big) 
		\frac{1}{|\mathcal{R}_{\ddot{n}_b}^{(p)}|}
		\underset{\ddot{n}_p \in \mathcal{R}_{\ddot{n}_b}^{(p)}}{\sum} Pr(\ddot{n}_b,\ddot{n}_p |\hat{\alpha}_b,\hat{\alpha}_p  )
		\end{align}
		Finally, by substituting the conditional path estimates $\hat{\alpha}_b  =\hat{\bm{V}}^{(b)}_{n_b,n_u}$ and $\hat{\alpha}_p =	( \bm{w}_c(|\ddot{n}_p|) )^H \bm{W}_c  \hat{\bm{V}}^{(p)}   \bm{F}_c^H
		\bm{a}_{U\!E} (  \phi^L_p(\ddot{n}_b,\ddot{n}_p,n_u) \big)$, and by approximating the estimation variance as being dominated by the \gls{AWGN} noise components, i.e., $\text{Var}[\hat{\alpha}_p]=\text{Var}[\hat{\alpha}_b]=N_0$, we can consider this probability across each of the equiprobable ranges $|{n}_b | =\ddot{n}_b  \in \{-n_b,n_b\}, |{n}_u | =\ddot{n}_u  \in \{-n_u,n_u\}$ to obtain
		\begin{align} \label{RAP_Prob_v_given_v}
		Pr(n_b,n_u| \hat{\bm{V}}^{(b)},\hat{\bm{V}}^{(p)}) 
		&=\frac{1}{4}  
		\sum_{ \substack{  
				\!\ddot{n}_b\in \{-n_b,n_b\} \\
				\ddot{n}_u\in \{-n_u,n_u\!\}}}
		Pr(\ddot{n}_b |\hat{\alpha}_b,\hat{\alpha}_p  ) \\
		&=\sum_{ \substack{  
				\!\ddot{n}_b\in \{-n_b,n_b\} \\
				\ddot{n}_u\in \{-n_u,n_u\!\}}}
		\underset{\ddot{n}_p \in \mathcal{R}_{\ddot{n}_b}^{(p)}}{\sum}
		\frac{1}{4|\mathcal{R}_{\ddot{n}_b}^{(p)}|}Pr(r_p(\ddot{n}_b,\ddot{n}_p)
		\Big| {\hat{\bm{V}}}^{(b)}_{n_b,n_u}) \times
		\nonumber \\
		&\qquad   Pr \left(r_p(\ddot{n}_b,\ddot{n}_p) \big|( \bm{w}_c(|\ddot{n}_p|) )^H  
		\bm{W}_c  \hat{\bm{V}}^{(p)}   \bm{F}_c^H
		\bm{a}_{U\!E} ( \phi^L_p(\ddot{n}_b,\ddot{n}_p,n_u)   \big)
		\right)
		\end{align}
		where $ \phi^L_p(\ddot{n}_b,\ddot{n}_p,n_u)$ can be found from (\ref{RAP_phi_L_p_cond}).
	\end{colar}
	\begin{exth}
		An example that illustrates conditional geometry is shown in Fig. \ref{RAP_RodFig}. Specifically, we show four conditional \gls{UE} positions and orientations, given the \gls{ROA} $\ddot{n}_b=3$ and $\ddot{n}_b=-3$ from the right-hand \gls{BS}. As is the case in our system model, the true \gls{UE} position is shown in the center (i.e., $\ddot{n}_b=-3$ in the example), with the correct \gls{ROD} directions shown as the beamforming directions with bold-solid lines (i.e., $\ddot{n}_u$). It follows that the correct directions will be expected to correspond to virtual channel entries that exhibit a strong path gain. The dashed and solid lines shown for the other conditional \gls{UE} positions represent the two possible \gls{UE} array orientations, each of which results in the considered correct beamforming vector (i.e., the expected strong estimate) directed back toward the right-hand \gls{BS} at (20,10). Focusing on the \gls{UE} in the center, it is notable that one of the two array orientations perfectly aligns the correct beamforming direction toward the left-hand \gls{BS} (i.e., it corresponds to the expected strong measurement in both virtual channels). As such, the probability of the conditional virtual channel entry will be high. 
		
		In contrast to the correct estimate, we can consider the alternative conditional \gls{UE} positions at coordinates (-20,-12.5) and (-18.5,12.5). At these positions, it is evident that neither of two conditional orientations direct the observed strong beamforming directions toward to the true \gls{BS} positions. As such, these candidate positions will yield low probabilities, as the expected measurements will not agree with the observed measurements. Focusing on the bottom-right \gls{UE} position, we see that one of the two orientations {\it{does}} result in the alignment of the strong \gls{UE} beamforming directions. However, as the resulting \gls{ROA} from the left-hand \gls{BS} is now incorrect, it will still correspond to a virtual channel entry with a weak gain and therefore result in low conditionally probability.
	\end{exth}
	
	\begin{note}
		It is worth noting that the geometric reasoning in Theorem 1 intentionally considers the scenario that applies to a joint \gls{LOS}---as asserted through Lemmas (\ref{RAP_RAPID_lem_b})-(\ref{RAP_RAPID_lem_d}). This set of conditions collectively considers geometric properties that would be consistent with a common line of sight path among two \gls{BS}. {In order to extend this framework to one that jointly estimates \gls{NLOS} paths, the developed model could also consider common scatterers alongside the already considered LOS components.} More specifically, for \gls{NLOS}, it may also be considered that, for the intercept of two \gls{ROA} pairs as a common \gls{NLOS} propagation source (i.e., a scatterer), the \gls{ROD} from the \gls{UE} must be the same, or very similar, for both \gls{BS} estimates. We have left this extension as future work.
	\end{note}
	
\end{thm}

\subsection{RAPID Summary}

Leveraging the expressions from the previous sub-sections, we now give a complete description our \gls{RAPID} beam training algorithm. We propose that after each \gls{BS} has collected \gls{UE} pilot symbols for $T_E$ channel estimation time slots, they each carry out their own channel estimation, before exchanging their estimates with nearby \gls{BS}. Initially, we assume that this information exchange is made possible by either a wired/wireless front-haul link between each \gls{BS}. We then propose a bandwidth limited exchange later in this sub-section.

Following (\ref{RAP_Prob_v_given_v}), after each \gls{BS} has exchanged its initial set of virtual channel estimates, the $b$th \gls{BS} can then compute its a priori virtual channel probabilities as
\begin{align} \label{RAP_Prob_v_given_v_avg}
Pr(n_b,n_u)=\frac{1}{B-1} \sum\limits_{
	\substack{\;p=1, \\p\neq b}}^B
Pr(n_b,n_u| \hat{\bm{V}}^{(b)},\hat{\bm{V}}^{(p)}).
\end{align}
With this result, each \gls{BS} can then select those \gls{UE}/\gls{BS} candidate beamforming pairs which have greatest probability of a path for data communication. The \gls{BS} can then feed back the selected \gls{UE} candidate beamforming indexes, requiring just $\text{log}_2(N_{})$ per index, for use in the following data communication\footnote{Alternatively, after the initial $T_E$ estimation time slots, the until-now transmitting \gls{UE} could instead start receiving with its continued pseudo-random beamforming sequence. As the \gls{UE} has been simultaneously associated with several \gls{BS}, all of which know this sequence and now have an estimate of which \gls{UE} beamforming directions are suitable for communication, this would only require, on average, $N_{\!U\!E}/(B \times R_{\!U\!E})$ time slots before a high path probability \gls{UE} beam direction is adopted. This opening could then be used to feed back the \gls{UE} side information and initiate communication.}.

The achievable link rate between the \gls{UE} and each \gls{BS} can then be expressed as \cite{rheath}
\begin{align} \label{RAP_Rate}
R_{opt}^{(u)}= \text{log}_2|\bm{I} + \frac{P}{N_0} \bm{W}_d^H \hat{\bm{H}}^{(b)} \bm{F}_d \bm{F}_d^H \hat{\bm{H}}^H \bm{W}_d|.
\end{align}
where $\bm{W}_d^H$ and $\bm{F}_d$ are \gls{BS} and \gls{UE} beamforming matrices consisting of the candidate beamforming vectors selected for communication. The remainder of this section proposes bandwidth-constrained information sharing and the application of \gls{RAPID} in downlink estimations.

\subsubsection{Bandwidth Constrained Ray Passing}

Due to the difficulties inherent in \gls{mmWave} communication, and the cost of wired back-haul in dense networks, it is possible that any communication channels between \gls{BS} links may be bandwidth-constrained. To reduce this overhead, we assume that each \gls{BS} is only able to share a limited number of entries from a virtual channel estimate. 

Fortunately, for any given \gls{BS} pair, $\hat{\bm{V}}^{(b)}$ and $\hat{\bm{V}}^{(p)}$, the complete set of entries is not needed, as they do not all have statistical dependencies. More specifically, as the two \gls{BS} exist in a 2D plane with \gls{ROA} bounded by positive radial distances, only half of the total number of $|\ddot{\mathcal{N}}_B|=2N_{\!B\!S}$ \gls{ROA} directions from one \gls{BS} have any directional component in relation to another \gls{BS}. As such, the largest number of \glspl{ROA} that can have a mutual intercept between two \gls{BS} is $(|\ddot{\mathcal{N}}_B|/2)^2 = N_{\!B\!S}^2$. Mathematically, we can denote the entries of the $b$th \gls{BS}'s virtual channel estimate that are passed to the $p$th \gls{BS} as
\begin{align} \label{RAP_H_pass}
\hat{\bm{V}}^{(b|p)}_{|\ddot{n}_b|,{n}_u} 
\Leftarrow 
\hat{\bm{V}}^{(b)}_{|\ddot{n}_b|,{n}_u} 
,   \forall  \; \; n_u, \ddot{n}_b| \exists \ddot{n}_b \in  \mathcal{R}_{\ddot{n}_p}^{(p)}, \forall \; \ddot{n}_p\in \ddot{\mathcal{N}}  
\end{align}
Returning to bandwidth constrained sharing, in some \gls{BS} deployments as little as half of these $N_{\!B\!S}^2$ \gls{ROA} intercepts correspond to unique entries in the virtual channel matrix. For example, when $\Theta_b=\Theta_p=0$ and both \gls{BS} are positioned on the x-axis, the \glspl{ROA} that are able to have intercepts are half in the positive \gls{AOA} range and half in the negative \gls{AOA}. Recalling the \gls{ULA} beam ambiguity problem (i.e., $\bm{a}_{B\!S}(\theta)=\bm{a}_{B\!S}(-\theta)$), the positive and negative angle ranges correspond to the same entries in the virtual channel matrix, due to the absolute index $|\ddot{\mathcal{N}}_B|$. In this case, only $N_{\!B\!S}N_{\!U\!E}/2$ entries need to be shared. Conversely, in other cases, such as when $\Theta_b=\Theta_p=0$ and both \gls{BS} are positioned on the y-axis, the angular range with a radial component toward the other \gls{BS} is either all positive or all negative, and thus all $N_{\!B\!S}N_{\!U\!E}$ entries are statistically relevant, to some extent. 

For very large \gls{MIMO} systems, this may still lead to an undesirable sharing overhead. Fortunately, owing to sparsity of the \gls{mmWave} channel, many of the estimated virtual channel entries are approximately zero and therefore can be neglected with little loss of performance. As such, we propose that, from the already reduced sets of virtual channel entries in (\ref{RAP_H_pass}), only the $N_d$ most dominant entries are shared. As the geometric representation of the \gls{mmWave} channel is inherently sparse, this has little effect on the performance of the system, provided $N_d$ is still greater than the number of paths. Furthermore, this also decreases computational complexity, as only \gls{ROA} pairs of significance need to be considered.

\begin{algorithm}
	
	\caption{\acrfull{RAPID} Beam Training.} \label{RAP_RAPID}
	{\fontsize{9}{11}\selectfont
		\vspace{0.15cm} $\mathbf{UE \; Input:}$ 
		Each \gls{UE} has a network-known candidate beamforming pilot sequence  $\sqrt{P} N_d \bm{s}_m \bm{F}_m, \forall\; m=1,\cdots, T_E$. 
		\vspace{0.15cm} \\
		
		$\mathbf{BS\;Input:}$ The orientation, $\Theta_p$, and relative position of nearby \gls{BS} $\bm{\Delta}_{b,p},\forall \;p =1,\cdots,B$.
		Each \gls{BS} has $N_d$, $N_0$, $\beta$, $A_g$ and knows $\bm{A}_m^{(b)}=( \bm{s}_m^T \bm{F}_m^T \bm{F}_c^*)  \otimes   (  (\bm{W}_m^{(b)})^H  \bm{W}_c  ) \forall\; m=1,\cdots, T_E. $ \vspace{0.15cm} \\
		
		$\mathbf{Initialization:}$ Each \gls{BS} pre-computes the \gls{ROA} intercepts \vspace{2.5pt} \\  
		$\;\mathcal{R}_{\ddot{n}_b}^{(p)} = \Big\{
		\ddot{n}_p \in \ddot{\mathcal{N}}_B \;\Big| 
		r_b(\ddot{n}_b,\ddot{n}_p)>0 \land
		r_p(\ddot{n}_b,\ddot{n}_p)>0\Big\}, \forall p,\ddot{n}_b$
		\vspace{0.15cm} \\
		
		\vspace{2.5pt} 
		$\mathbf{Transmission\;and\;Independent\;Estimation:}$ \vspace{2.5pt} \\
		
		\For( \emph{}){$m=1,2,\cdots,T_E$}
		{
			\vspace{2.5pt} 
			// \gls{UE} Transmits beamformed pilots \vspace{2.5pt} \\
			$\bm{x}_m = \sqrt{\frac{P}{R_{\!U\!E}}} \bm{F}_m \bm{s}_m$     \vspace{5.0pt} \\
			
			\For( \emph{}){$b=1,2,\cdots,B$}{
				\vspace{2.5pt} 
				// The $b$th \gls{BS} receives with $\bm{W}_m$ to obtain \vspace{2.5pt} \\
				$\bm{y}^{(b)}_m = (\bm{W}_m^{(b)})^H(  \bm{H}^{(b)}\bm{x}_m + \bm{q}_m^{(b)})$ \vspace{5.0pt} \\
				
			}
			
		} \vspace{2.5pt}

		\For( \emph{}){$b=1,2,\cdots,B$}{
			// The $b$th \gls{BS} uses $\bm{y}^{(b)} = [\bm{y}^{(b)}_1;\cdots;\bm{y}^{(b)}_{T_E}]$ and \\ $\bm{A}^{(b)}=[\bm{A}_1^{(b)};\cdots;\bm{A}_{T_E}^{(b)}]$ for independent sparse recovery: \vspace{2.5pt} \\
			$\;\;\;\;\hat{\bm{v}}^{(b)} = \underset{\bm{v}}{\operatorname{argmin}}\Big[ || \bm{y}^{(b)} -  A_g \bm{A}^{(b)} \bm{v}||_2^2 + \gamma || \bm{v} ||_1 \Big]$
			\vspace{5.0pt} \\
		}
		\vspace{2.5pt} 
		$\mathbf{ Ray\; Passing\; and\;Indirect\;Estimation:}$  \vspace{2.5pt} \\
		\For( \emph{}){$b=1,2,\cdots,B$}{
			\For( \emph{}){$p=1,2,\cdots,B, |\; p\neq b$ \vspace{2.5pt}}{
				
				// The $b$th \gls{BS} passes its $N_d$ strongest entries to the $p$th \gls{BS}, which has a common \gls{ROA} intercept as \vspace{4.5pt} \\
				$
				\hat{\bm{V}}^{(b|p)}_{|\ddot{n}_b|,{n}_u} 
				\Leftarrow
				\hat{\bm{V}}^{(b)}_{|\ddot{n}_b|,{n}_u} 
				, \forall  \; \; n_u, \ddot{n}_b| \exists \ddot{n}_b \in  \mathcal{R}_{\ddot{n}_p}^{(p)}, \forall \; \ddot{n}_p\in \ddot{\mathcal{N}}  $
				\vspace{2.5pt} \\ 
				
				// The $b$th \gls{BS} receives $N_d$ entries from the $p$th \gls{BS} as \vspace{4.5pt} \\ 
				
				$\tilde{\bm{V}}^{(p)}_{|\ddot{n}_p|,{n}_u}   
				\Leftarrow
				\hat{\bm{V}}^{(p|b)}_{|\ddot{n}_p|,{n}_u}
				,   \forall  \; n_u, \ddot{n}_p |\exists \ddot{n}_p  \in  \mathcal{R}_{\ddot{n}_b}^{(b)},   \forall  \;\ddot{n}_b \in \ddot{\mathcal{N}} 	$
				\vspace{8.5pt} \\ 
				
				// The $b$th \gls{BS} computes its the conditional probability from (\ref{RAP_Prob_v_given_v})  
				\vspace{4.5pt} \\ 
				$Pr(n_b,n_u| \hat{\bm{V}}^{(b)},\tilde{\bm{V}}^{(p)}) 
				=\frac{1}{4}  
				\sum_{ \substack{  
						\!\ddot{n}_b\in \{-n_b,n_b\} \\
						\ddot{n}_u\in \{-n_u,n_u\!\}}}
				Pr(\ddot{n}_b |\hat{\alpha}_b,\hat{\alpha}_p  ) , \forall n_b,n_u $
				\vspace{4.5pt} \\ 	
			}
			// Compute conditional probability given all other \gls{BS}. 
			\vspace{4.5pt} \\ 
			$Pr(n_b,n_u) 
			=\frac{1}{4}  
			\sum_{ \substack{  
					p=1 \\
					p\neq b  }}^B
			Pr(n_b,n_u| \hat{\bm{V}}^{(b)},\tilde{\bm{V}}^{(p)}), \forall n_b,n_u $ \\
		}
		$\mathbf{Output:}$ $Pr(n_b,n_u)\forall n_b,n_u$ .
	}
\end{algorithm}

Using this approach, we denote the constrained matrix received by the $b$th \gls{BS} from the $p$th \gls{BS} as $\tilde{\bm{V}}^{(p)}$, such that $||\tilde{\bm{V}}^{(p)}||_0=N_d$ as
\begin{align} \label{RAP_H_pass_2}
\tilde{\bm{V}}^{(p)}_{|\ddot{n}_p|,{n}_u}   
\Leftarrow
\tilde{\bm{V}}^{(p|b)}_{|\ddot{n}_p|,{n}_u}
,   \forall  \; n_u, \ddot{n}_p |\exists \ddot{n}_p  \in  \mathcal{R}_{\ddot{n}_b}^{(b)},   \forall  \;\ddot{n}_b \in \ddot{\mathcal{N}}
\end{align}
With this constrained information, we can rewrite (\ref{RAP_Prob_v_given_v_avg}) as

\begin{align} \label{RAP_Prob_v_given_v_avg_2}
Pr(n_b,n_u)=\frac{1}{B-1} \sum\limits_{
	\substack{\;p=1, \\p\neq b}}^B
Pr(n_b,n_u| \hat{\bm{V}}^{(b)},\tilde{\bm{V}}^{(p)}).
\end{align}
We show the complete \gls{RAPID} beam training approach in Algorithm \ref{RAP_RAPID}.

\begin{figure*}[!p]
	\centering
	\subfigure[][]{\includegraphics[width=4.5in,trim={0.0cm 0.0cm 0.0cm 0.0cm},clip]{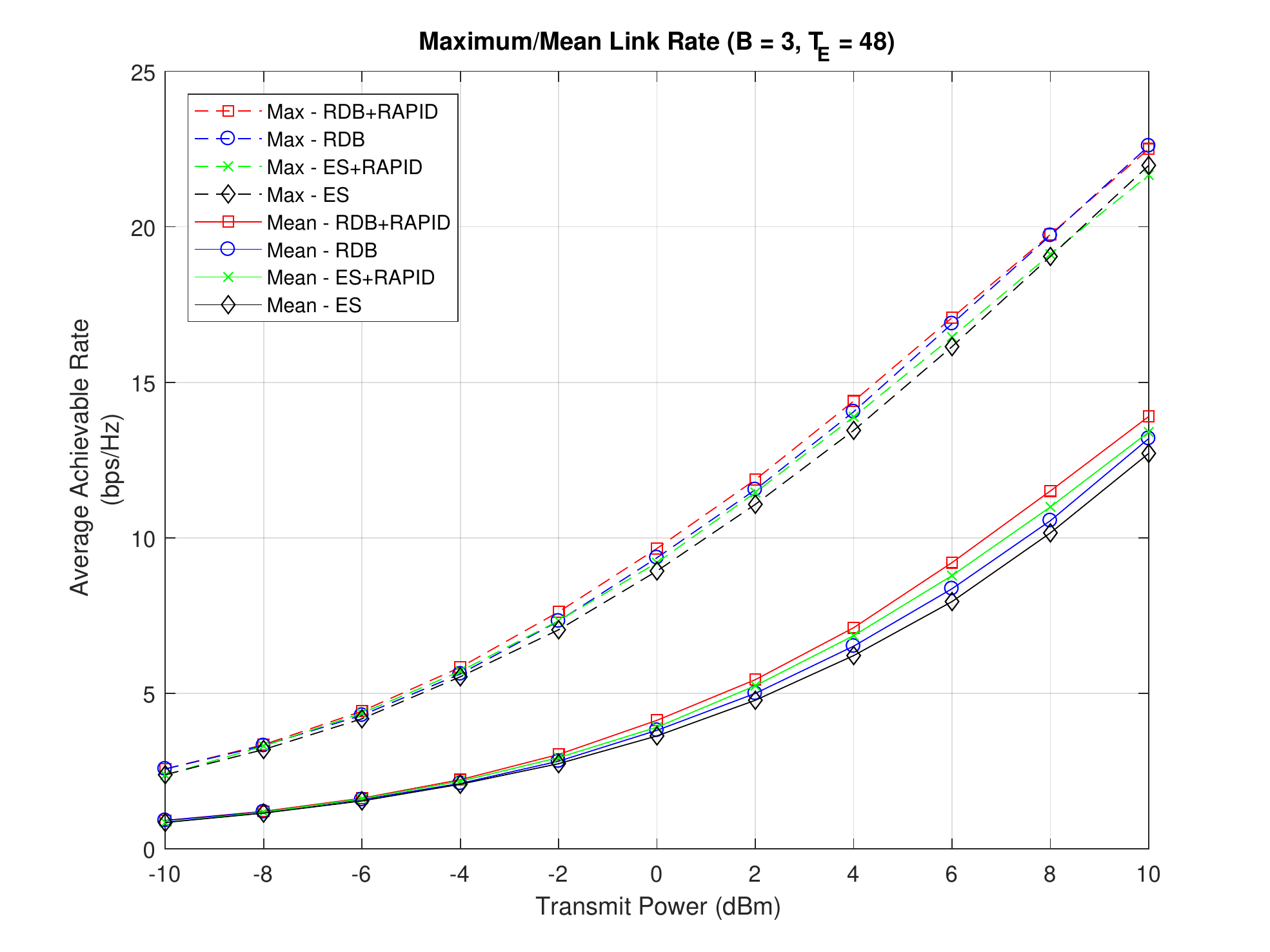}}
	\subfigure[][]{\includegraphics[width=4.5in,trim={0.0cm 0.0cm 0.0cm 0.0cm},clip]{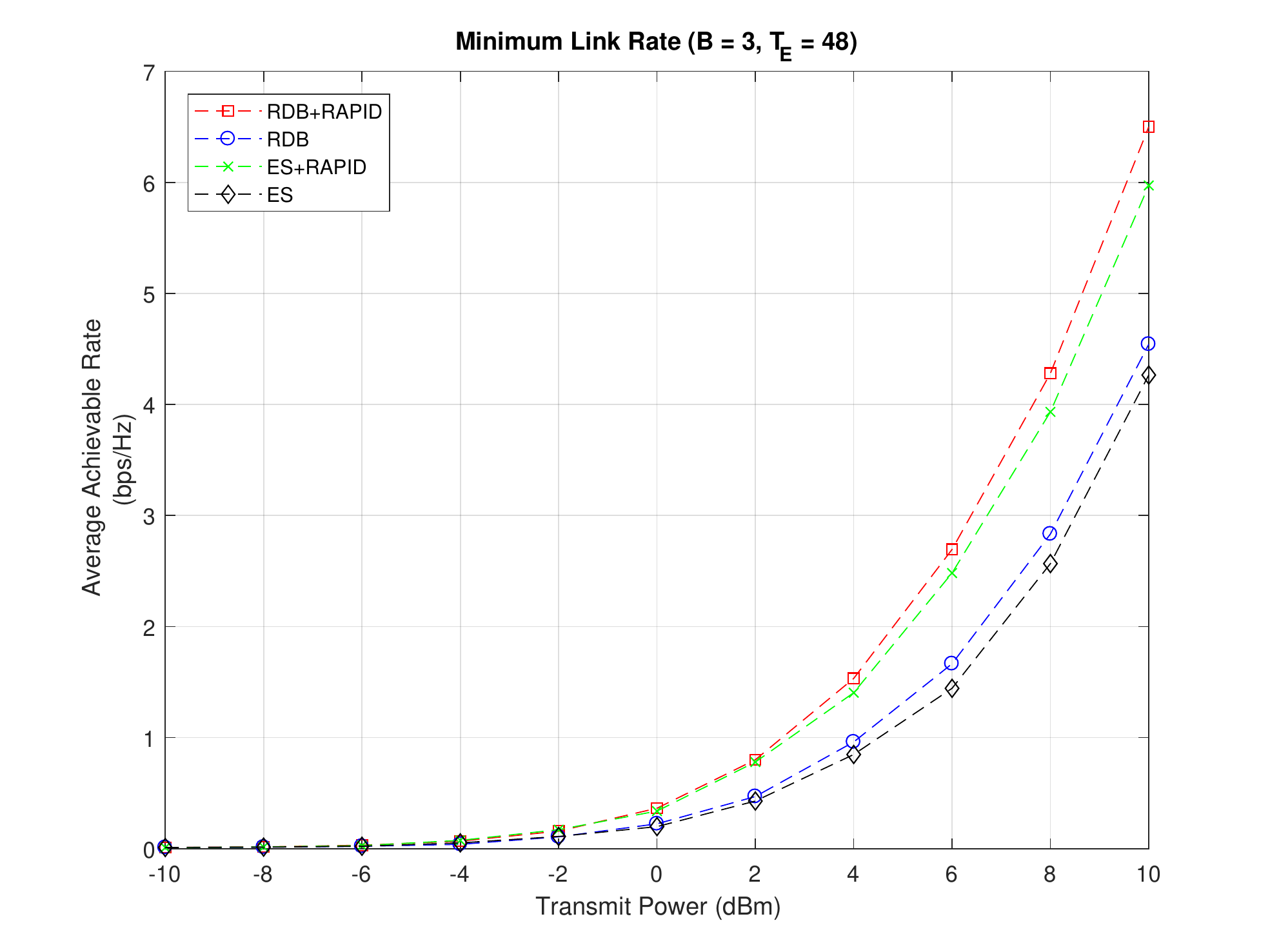}}
	\caption[Rate results where the network consists of $B=3$  \acrfull{BS} each equipped with $N_{\!B\!S}=32$ antenna and $R_{\!B\!S}=8$  \acrfull{RF} chains and the user is equipped with $N_{\!U\!E}=16$ antenna and $R_{\!U\!E}=4$ {RF} chains.]
	{ results where the network consists of $B=3$  \acrfull{BS}, each equipped with $N_{\!B\!S}=32$ antennas and $R_{\!B\!S}=8$  \acrfull{RF} chains and the user is equipped with $N_{\!U\!E}=16$ antennas and $R_{\!U\!E}=4$ \gls{RF} chains. We assume the expected number of paths is $E[L]=3$. (a) Shows the average maximum and mean achievable link rates after estimation, while (b) shows the minimum achievable link rate.}
	\label{RAP_B_3_res_rate}
\end{figure*}

\begin{figure*}[!p]
	\centering	
	\subfigure[][]{\includegraphics[width=4.5in,trim={0.0cm 0.0cm 0.0cm 0.0cm},clip]{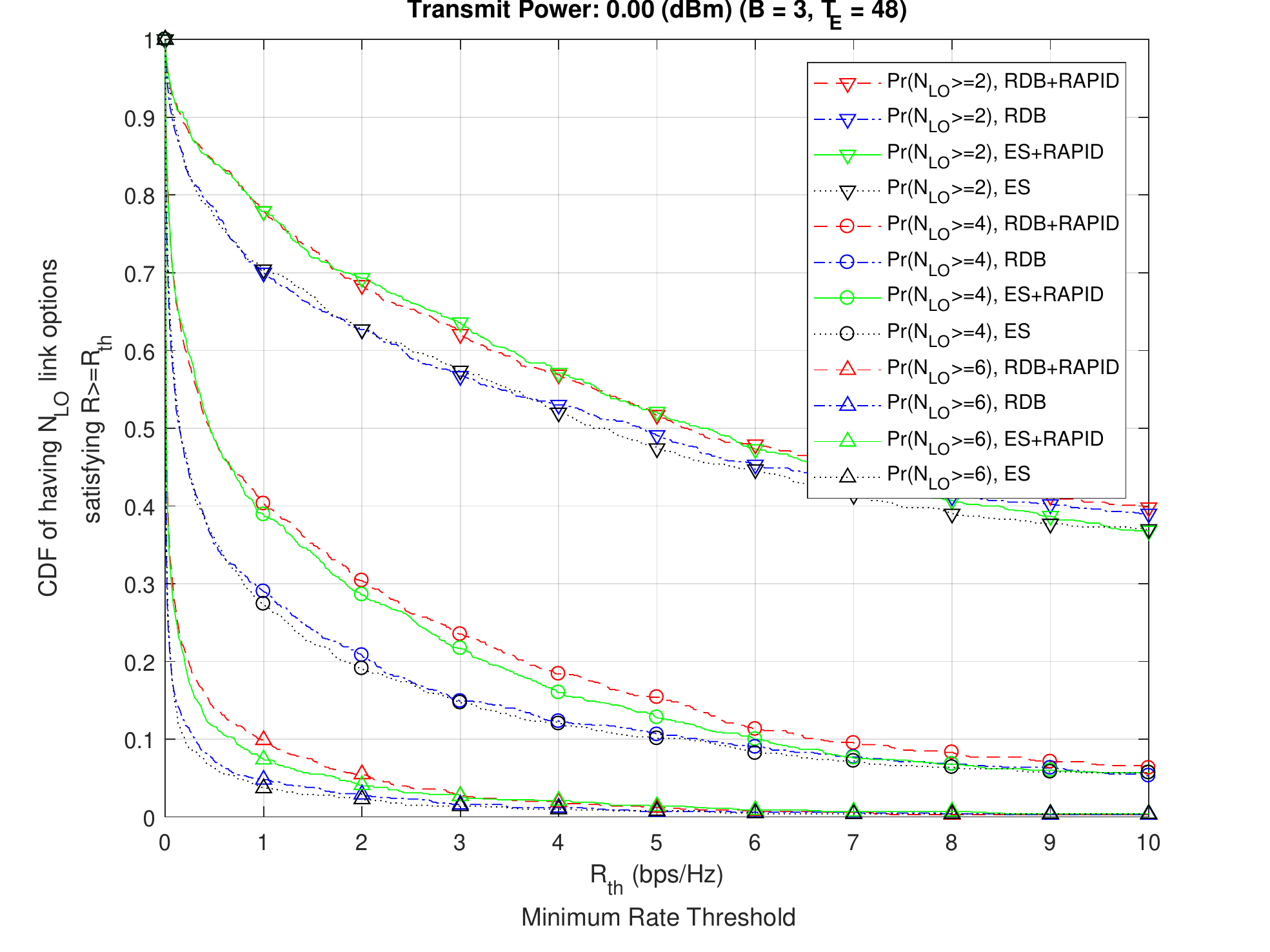}}
	\subfigure[][]{\includegraphics[width=4.5in,trim={0.0cm 0.0cm 0.0cm 0.0cm},clip]{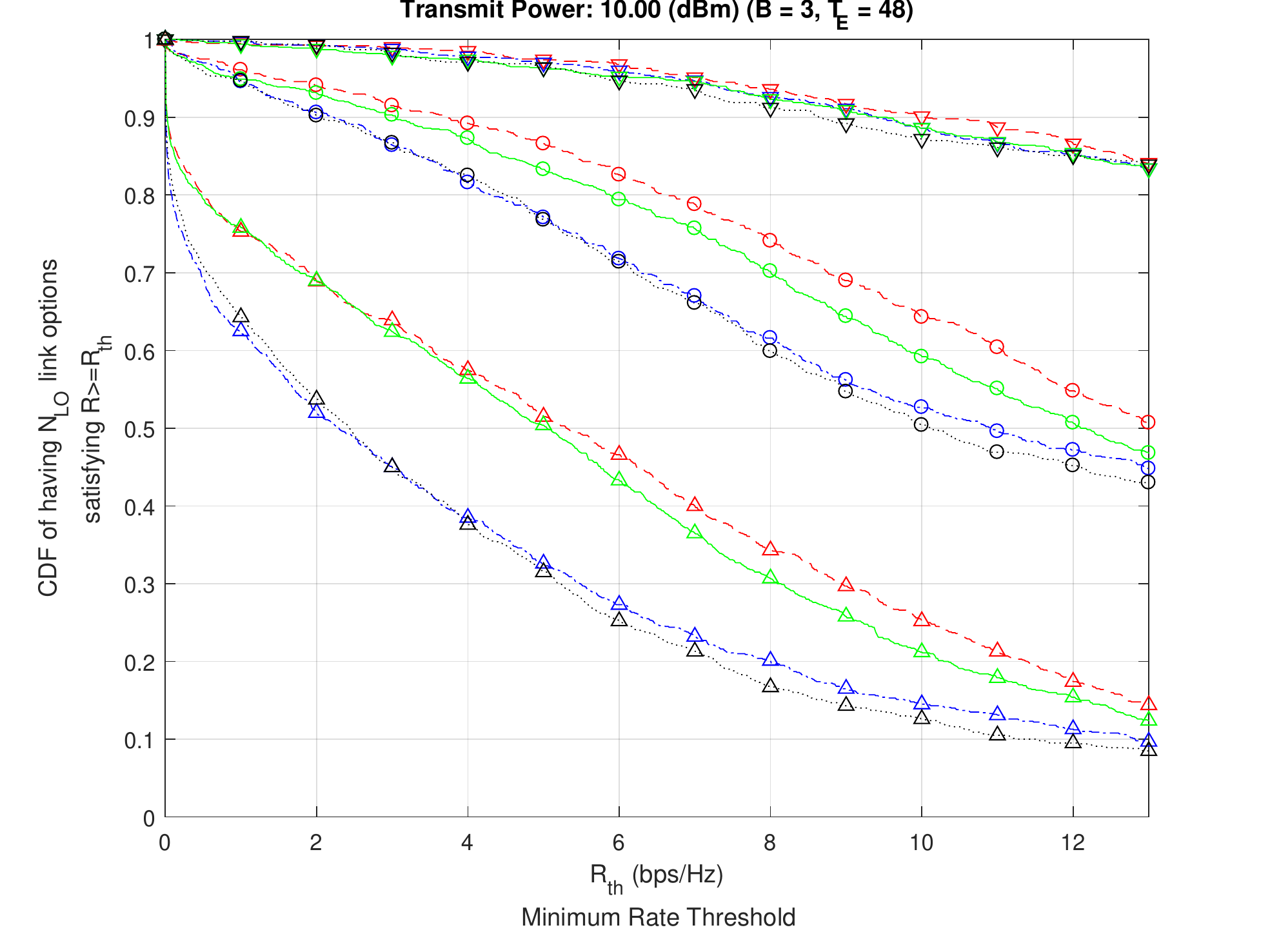}}
	\caption[Coverage \acrfull{CDF} for a network consisting of $B=3$  \acrfull{BS}, each equipped with $N_{\!B\!S}=32$ antennas and $R_{\!B\!S}=8$  \acrfull{RF} chains and the user is equipped with $N_{\!U\!E}=16$ antennas and $R_{\!U\!E}=4$ {RF} chains.]
	{Numerical results where the network consists of $B=3$  \acrfull{BS}, each equipped with $N_{\!B\!S}=32$ antennas and $R_{\!B\!S}=8$  \acrfull{RF} chains and the user is equipped with $N_{\!U\!E}=16$ antennas and $R_{\!U\!E}=4$ \gls{RF} chains. We assume the expected number of paths is $E[L]=3$. (c) shows the \acrfull{CDF} of network coverage for $P=0$ dBm and (d) $P=10$ dBm.}
	\label{RAP_B_3_res_cov}
\end{figure*}

\subsubsection{RAPID Downlink Beam Training}

Up to this point, we have introduced \gls{RAPID} as a cooperative uplink channel estimation strategy; however, by considering that the only prior knowledge required to compute (\ref{RAP_Prob_v_given_v_avg}) is the relative positions of each network \gls{BS} and their orientations, \gls{RAPID} can also be implemented in downlink at the \gls{UE}. To this end, the \gls{UE} would only require this static network deployment information, along with a coarse estimate its network position, in order to reduce the number of considered \gls{BS}. Then, similarly to the uplink description, each \gls{BS} can broadcast beamformed pilot signals with orthogonal spreading codes while each \gls{UE} collects the signals with a sequence of beamforming directions. After the $T_E$ estimation time slots, each \gls{UE} can make a direct estimate of the downlink channel and implement \gls{RAPID} with no communication overhead for the network. 

In this scenario, although there is no sharing overhead, the computational burden that was originally in the up-link and distributed among several \gls{BS} would now need to be carried out by a single \gls{UE}. As the \gls{UE} is expected to have less computational power and more stringent energy requirements, it may still be beneficial for the \gls{UE} to only consider the $N_d$ most dominant and dependent entries from each estimate. This downlink estimation strategy would also still permit the \gls{UE} to adapt its beamforming directions during the estimation process, as proposed in \cite{kokshoorn2017beam}.  

\section{Numerical Results}

We now provide some numerical results to evaluate the performance of our proposed scheme. We consider an \gls{mmWave} system where a \gls{UE} is equipped with $N_{\!U\!E} = 16$ antennas and $R_{\!U\!E}=4$ \gls{RF} chains. We assume this \gls{UE} to be within the range of a network of $B$ \gls{BS}, each equipped with $N_{\!B\!S} = 32$ antennas and $R_{\!B\!S}=8$ \gls{RF} chains. We adopt a user-centric deployment network model in which each \gls{BS} is positioned within a $100$m x $100$m grid, with the \gls{UE} at its center (i.e., a maximum of $50$m away from the \gls{UE} along the x- or y-axis). We consider each \gls{BS} to follow a uniform random distribution within this space, while the orientation of each \gls{BS} also follows a uniformly random distribution in the continuous range $[0,2\pi]$. Similarly, we consider \gls{UE} orientation to also follow a uniform random distribution in the range $[0,2\pi]$. The resulting deployment dependent channels can therefore be found from (\ref{RAP_H_G}), in which the path loss exponent is considered as $\beta=4$ to represent severe \gls{mmWave} propagation losses. We consider each receiver's noise power to be $N_0=10^{-5}$, such that the propagation path \gls{SNR} can be expressed as $\sigma_R^2/N_0$, which leads to a minimum link \gls{SNR} of  $r_b^{-\beta}/N_0=-24$ dB at $\text{max}[r_b]=\sqrt{2} \times 50m$.

\begin{figure*}[!p]
	\centering	
	\subfigure[][]{\includegraphics[width=4.5in,trim={0.0cm 0.0cm 0.0cm 0.0cm},clip]{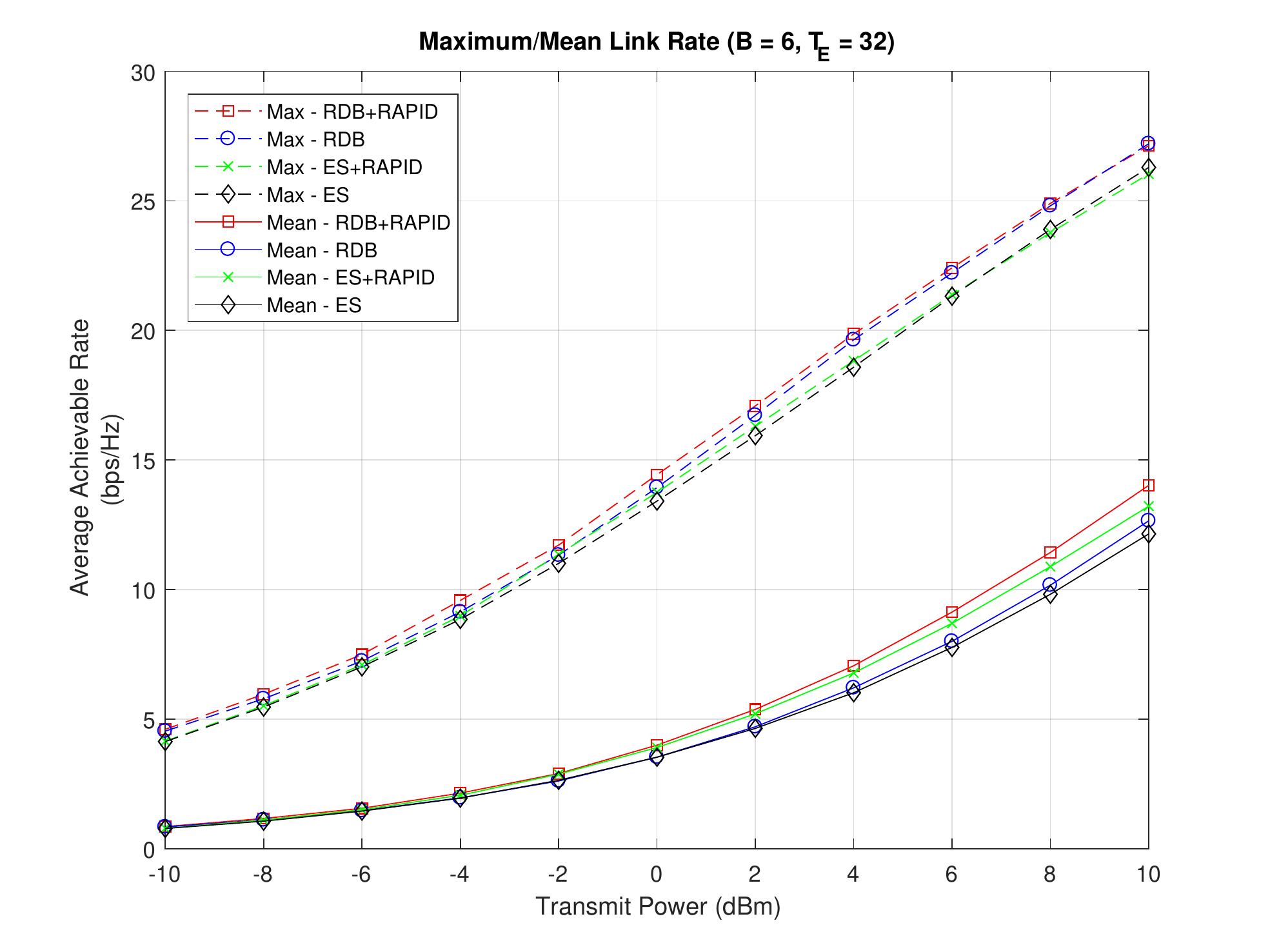}}
	\subfigure[][]{\includegraphics[width=4.5in,trim={0.0cm 0.0cm 0.0cm 0.0cm},clip]{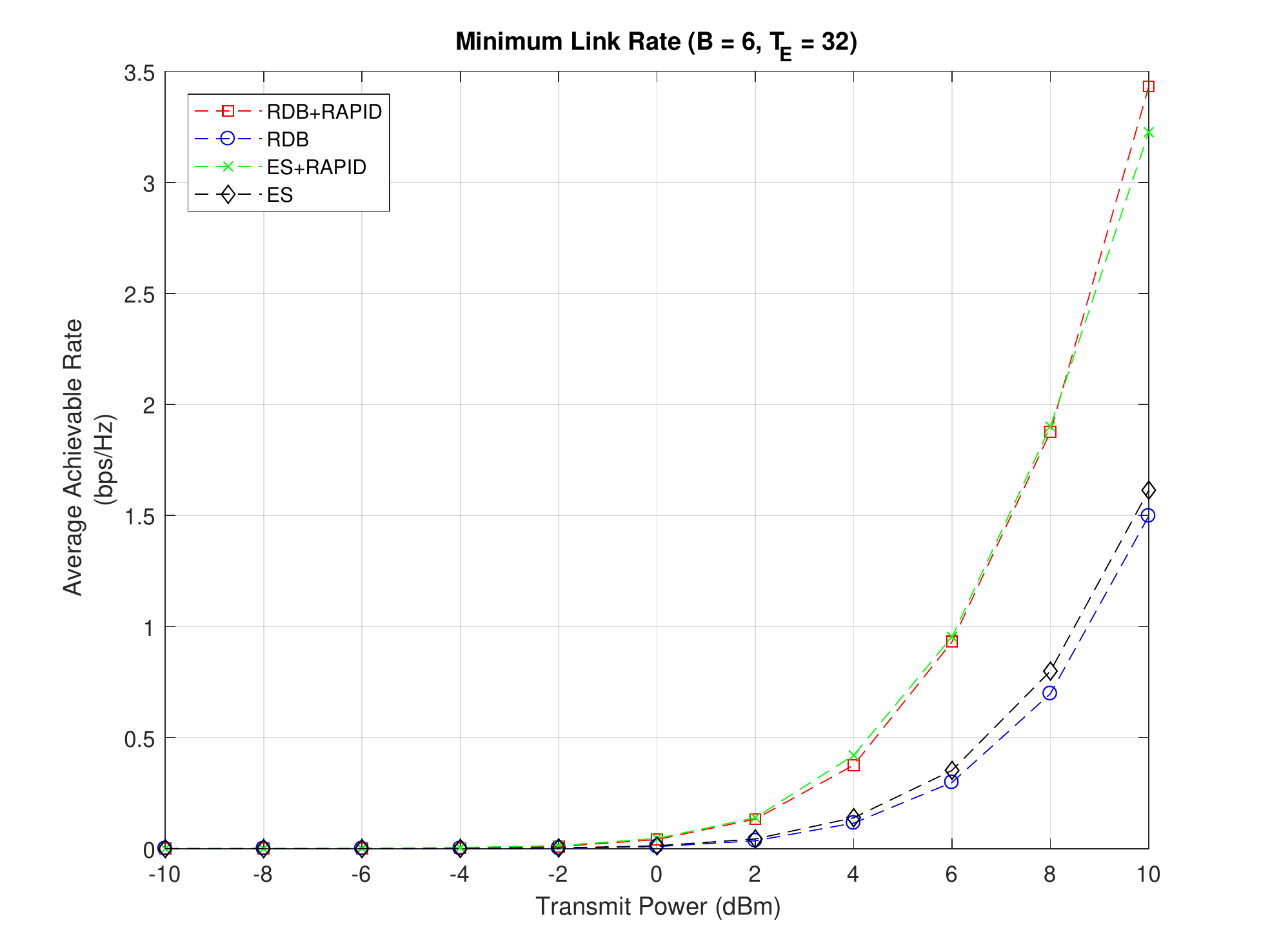}}
	\caption[Rate results where the network consists of $B=6$  \acrfull{BS}, each equipped with $N_{\!B\!S}=32$ antennas and $R_{\!B\!S}=8$  \acrfull{RF} chains and the user is equipped with $N_{\!U\!E}=16$ antennas and $R_{\!U\!E}=4$ {RF} chains.]
	{Numerical results where the network consists of $B=6$  \acrfull{BS}, each equipped with $N_{\!B\!S}=32$ antennas and $R_{\!B\!S}=8$  \acrfull{RF} chains and the user is equipped with $N_{\!U\!E}=16$ antennas and $R_{\!U\!E}=4$ {RF} chains. (a) Shows the average maximum and mean achievable link rates after estimation, while (b) shows the minimum achievable link rate.}
	\label{RAP_B_6_res_rate}
\end{figure*}

\begin{figure*}[!p]
	\centering
	\subfigure[][]{\includegraphics[width=4.5in,trim={0.0cm 0.0cm 0.0cm 0.0cm},clip]{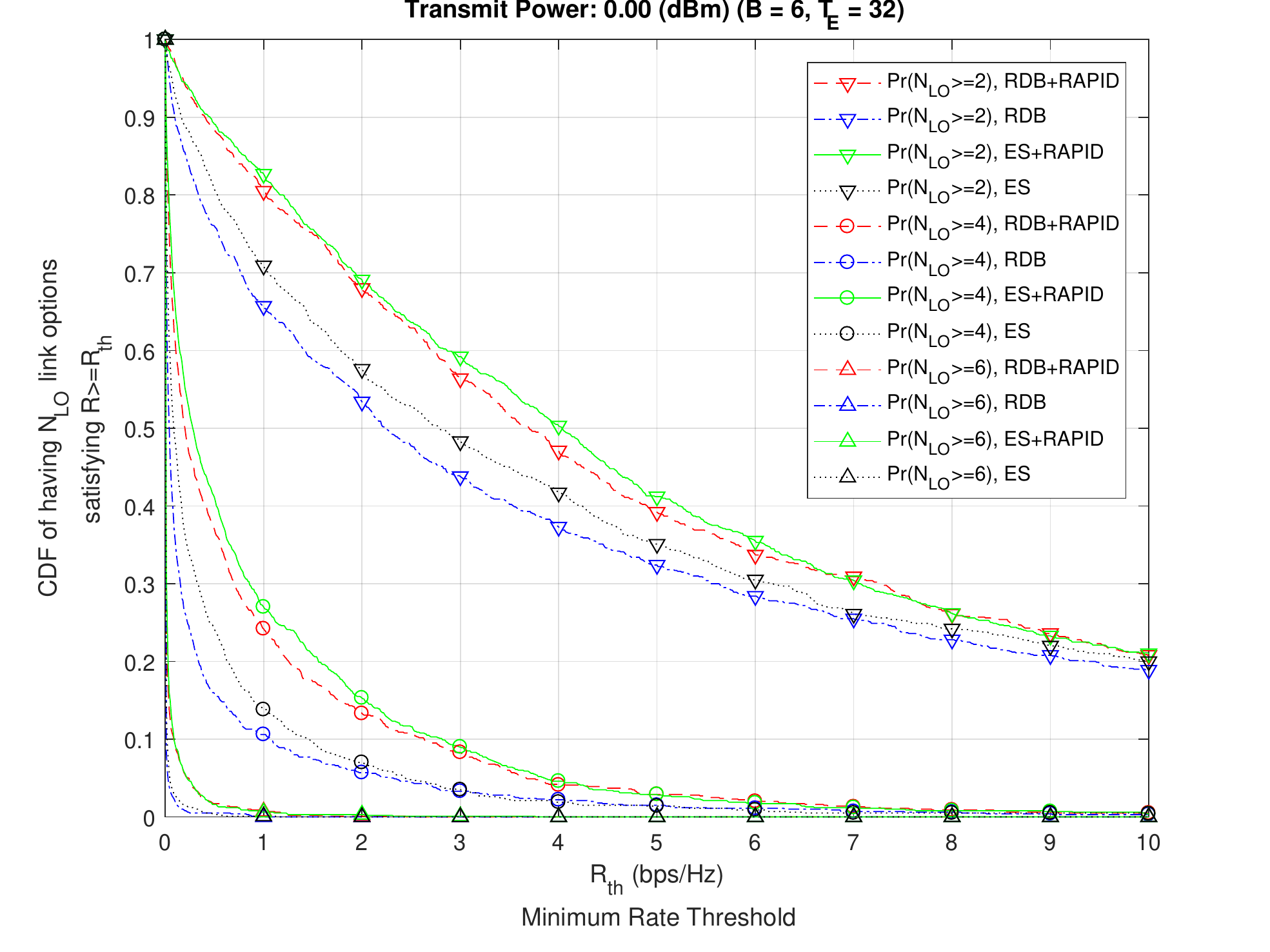}}
	\subfigure[][]{\includegraphics[width=4.5in,trim={0.0cm 0.0cm 0.0cm 0.0cm},clip]{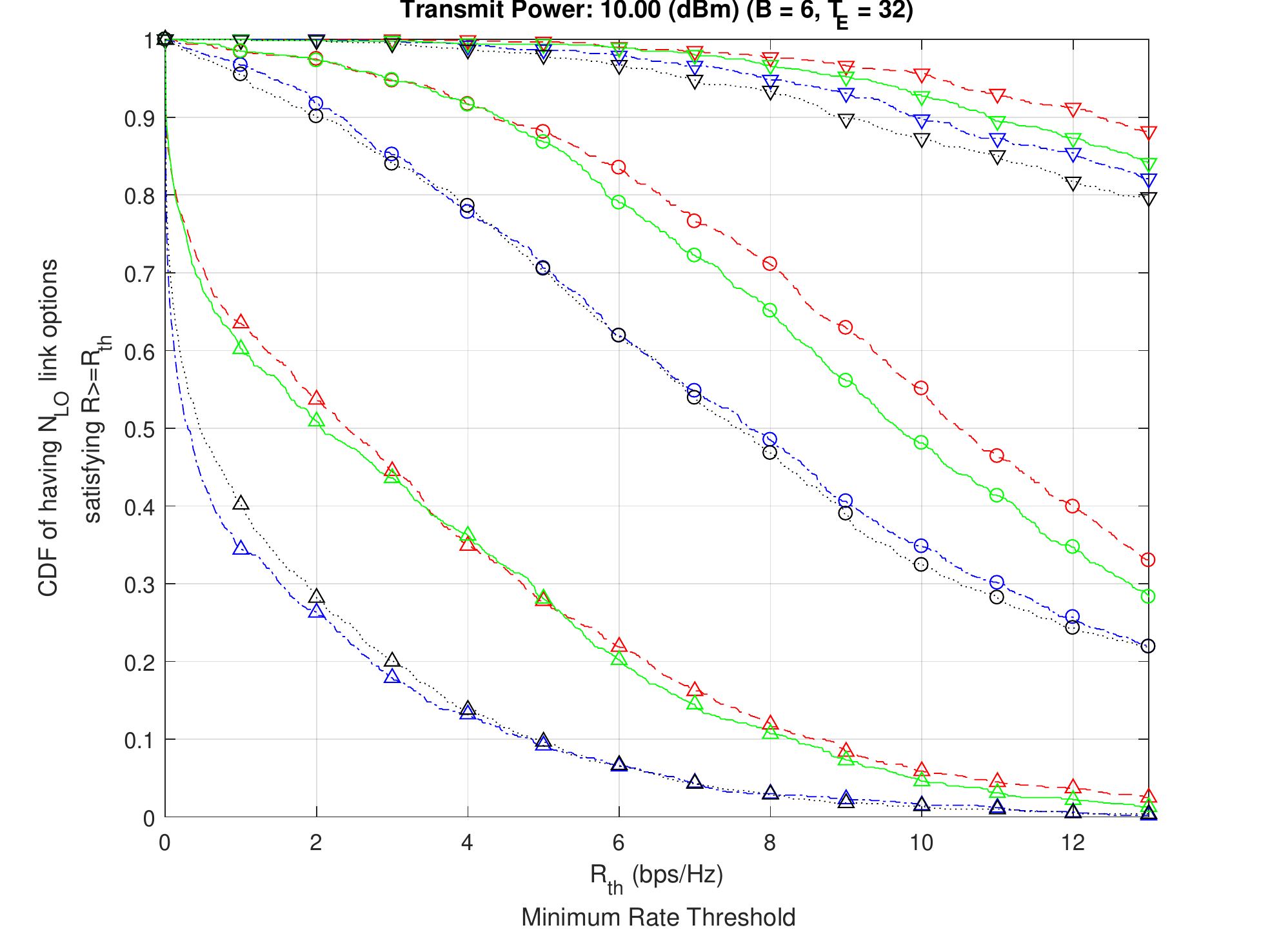}}
	\caption[Coverage \acrfull{CDF} for a network consisting of $B=6$  \acrfull{BS}, each equipped with $N_{\!B\!S}=32$ antennas and $R_{\!B\!S}=8$  \acrfull{RF} chains and the user is equipped with $N_{\!U\!E}=16$ antennas and $R_{\!U\!E}=4$ {RF} chains.]
	{Numerical results where the network consists of $B=6$  \acrfull{BS}, each equipped with $N_{\!B\!S}=32$ antennas and $R_{\!B\!S}=8$  \acrfull{RF} chains and the user is equipped with $N_{\!U\!E}=16$ antennas and $R_{\!U\!E}=4$ {RF} chains. (c)  shows the  \acrfull{CDF} of network coverage for $P=0$ dBm and (d)  $P=10$ dBm.}
	\label{RAP_B_6_res_cov}
\end{figure*}

We apply \gls{RAPID} to \gls{RDB} with $T_E$ channel estimation time slots and an \gls{ES}-based channel estimation with $T_{ES}=N_{\!U\!E}N_{\!B\!S}/R_{\!B\!S}=64$ estimation time slots. To compare the performance of each scheme as a result of estimation, we show the average best-available link rate (i.e., the maximum achievable rate given all estimated channels) along with the average of the worst available link rate (i.e., the minimum achievable link rate given all estimated channels). For completeness, we also include the average of all available links. To demonstrate coverage probability and link redundancy, we also show the \gls{CDF} of the \gls{UE} with $N_{LO}$ link options, whereby we impose the requirement that a link must satisfy $R>R_{th}$ along with a coverage rate threshold $R_{th}$.

In Fig. \ref{RAP_B_3_res_rate} (a), we show both the maximum achievable link rate and the average achievable link rate for a network of $B=3$ \gls{BS}, with \gls{RDB} using $T_E=48$. In most cases, it is evident that \gls{RDB} tends to outperform \gls{ES} despite using fewer measurement timeslots, particularly at a high \gls{SNR}, because \gls{RDB} has greater measurement diversity due to pilots being transmitted with multiple beamforming directions in each time slot. This effectively allows the receiver to sample several entries of the virtual channel at once. Conversely, \gls{ES} sequentially transmits a pilot with only one beamforming direction at a time and is included as a benchmark approach. By comparing the average link rate of the schemes in Fig. \ref{RAP_B_3_res_rate} (a), it is notable that those using \gls{RAPID} show little advantage at a low \gls{SNR}; however, as transmit power increases, both \gls{ES} and \gls{RDB} are able to achieve an average link rate increase of around 1 bit/s/Hz. Interestingly, comparing this to maximum link rate performance, we find that the there is very little increase as \gls{SNR} increases, because---in most cases---the \gls{BS} that has the best channel in relation to the \gls{UE} is the one that stands to gain the least from sharing its information with the other \gls{BS}. Conversely, in Fig. \ref{RAP_B_3_res_rate} (b), we show the average of the minimum achievable link rate, which effectively represents the \gls{BS} that has the worst channel in relation to the \gls{UE}, which is therefore the \gls{BS} that stands to gain the most from exchanging information. As the \gls{UE} transmit power increases, we can see that the minimum link rate of the systems using \gls{RAPID} increase significantly by up to around 2bps/Hz at a transmit power of 10 dBm.

Turning our attention to Figs. \ref{RAP_B_3_res_cov} (a) and (b), we show the \glspl{CDF} for a number of achievable link options, for $P=0$ dBm and $P=10$ dBm, respectively. In both cases, we can see that \gls{RAPID} is able to increase significantly the probability of having a larger number of available link options, particularly for lower-rate requirement thresholds. This is an inherent property of \gls{RAPID}'s ability to improve significantly the weaker network links. This low rate threshold region also fits for \gls{mmWave} systems, as throughput gains are expected to come from large bandwidths as opposed to complex modulation schemes. For much greater rate thresholds we see that the available link probabilities of all systems tend to converge. 

In Fig. \ref{RAP_B_6_res_rate} and Fig. \ref{RAP_B_6_res_cov}, we increase \gls{BS} density to $B=6$ for the same deployment area. We also reduce the number of \gls{RDB} time slots to $T_E=32$. In (a) and (b), we again show the minimum, mean, and maximum link rates. Again, we can see from (b) that the minimum link rate is able to increase by around 2 bits/s/Hz by applying \gls{RAPID} despite the worst of the $B=6$ channels being much worse than $B=3$, as in Fig. \ref{RAP_B_3_res_rate}. Looking at the average link rate in Fig. \ref{RAP_B_6_res_rate} (b), we can see a more noticeable increase with $B=6$ for the same reason, as now there are many more \gls{BS} to benefit from the the better channels that are shared. Turning to the \glspl{CDF} in \ref{RAP_B_6_res_cov} (a) and (b), we find that the probability of having more available links is still much greater with \gls{RAPID}, in particular at a high \gls{SNR}. 

\section{Conclusion}

In this paper we proposed have a cooperative \gls{mmWave} beam training scheme in which multiple network BS share information to enhance the channel estimation accuracy of one another---and therefore the network performance as whole. In order to combine shared information, we proposed a \acrfull{RAPID} algorithm in which the probability of each directional path can be conditionally considered by multiple BS. By leveraging the derived statistical relations, it was proposed that each BS need only share information with other BS that are able to utilize it, thus reducing the communication overhead involved in this information sharing. The presented results established that the BS link which has the worst quality benefits the most from the scheme. Furthermore, by considering a minimum rate threshold for communication, we demonstrated that RAPID is able to increase significantly the probability that one or more links are available to a user at any given time.

\bibliographystyle{IEEEtran}
\bibliography{library}

\end{document}